\documentclass[aps,prx,twocolumn,superscriptaddress,showpacs]{revtex4-1}
\usepackage{hyperref}
\usepackage{times,xspace}
\usepackage{graphicx,graphics,color,epsfig}
\usepackage{amsmath}
\usepackage{amssymb}
\usepackage{amsfonts}
\usepackage{amsfonts}
\usepackage{epstopdf}
\usepackage{bm}
\usepackage{hyperref}
\usepackage{longtable}     
\usepackage{url}           
\usepackage{bm}            
\usepackage{color}
\usepackage{float}
\usepackage[normalem]{ulem}

\begin{document}

\title{Nodeless superconducting gap induced by odd parity pair density wave in underdoped cuprates}

\author{Tanmoy Das\\
\normalsize{Department of Physics, Indian Institute of Science, Bangalore 560012, India.}}

\date{\today}

\begin{abstract}
We present a theoretical model to show that the transition from an antiferromagnetic (AFM) phase to the $d_{x^2-y^2}$-wave superconductivity occurs through a robust companion of a finite momentum pairing between $({\bf k},\uparrow)$ and $(-{\bf k}-{\bf Q},\downarrow)$ electrons, namely pair-density wave (PDW) state, where ${\bf Q}=(\pi,\pi)$ is the AFM wavevector. Interestingly, the spatial structure of the PDW is constrained to be a $p+ip$-wave symmetry, which follows $p_{\bf k}=p_{-{\bf k}-{\bf Q}}$ under fermions exchange in cuprates, dictating a spin-singlet pairing. Furthermore, the PDW state produces a fully gapped quasiparticle spectrum which explains recent observations of the fully gapped quasiparticle structure of lightly doped cuprates in the hole doping side. We study the stability of all three phases within the self-consistent mean-field theory. Finally, we calculate the superfluid density to propose its exponential temperature dependence as a test to the SC origin of this fully gapped state, while the phase modulation of the PDW state can be visualized by scanning probes.
\end{abstract}



\maketitle

\section{Introduction}

A fascinating fact of cuprate research is that despite three decades of extensive research, experiments still uncover new and unexpected results which surprise our hitherto reached consensus. A latest addition to this investigation comes from various spectroscopic findings of a fully gapped electronic structure in deeply underdoped region,\cite{LSCO2000,ccoc,bi22122006,bi2212,LSCO2013,bi2201,YBCO} where nodal $d$-wave pairing symmetry is expected. This feature is ubiquitously established in many hole-doped cuprates studied so far, including La-based,\cite{LSCO2000,LSCO2013} Bi-based,\cite{bi22122006,bi2212,bi2201} Cl-based,\cite{ccoc} and also Yb-based compounds.\cite{YBCO} This unexpected result led to questions such as: Is the fully gapped state related to superconductivity other than the $d$-wave, or is it derived from the exotic normal state, or does it arise from the interplay of superconductivity with any normal state phase, or others?

Theory of nodeless $d$-wave pairing state is demonstrated earlier to exist in electron-doped cuprates,\cite{TanmoyED,CSTing} and also in 122 iron-chalcognenides.\cite{TanmoyFeSe,NodelessFeSe1,NodelessFeSe2} However, in these cases either interaction\cite{TanmoyED,CSTing} or band structure topology\cite{TanmoyFeSe,NodelessFeSe1,NodelessFeSe2} removes the nodal states in the non-SC state itself, which thus yields a fully gapped electronic structure despite $d$-wave being the underlying pairing symmetry. But this description does not apply to the nodeless state in the hole-doped cuprates since as temperature is raised above the SC state, a truncated Fermi surface (FS) is experimentally verified to exist in the nodal region.\cite{LSCO2013} Motivated by this unexpected behavior, several possible explanations have been put forward in recent times which include Coulomb disorder effect,\cite{CoulombGap} polaron effect,\cite{bi2201} $d+is$ pairing,\cite{Annica} topological superconductivity,\cite{TSC}, dynamical mean-field theory calculations\cite{CDMFT1,CDMFT2}. However, no consensus is yet reached.
 
\begin{figure}[ht]
\hskip-0.6cm
\rotatebox[origin=c]{0}{\includegraphics[width=0.99\columnwidth]{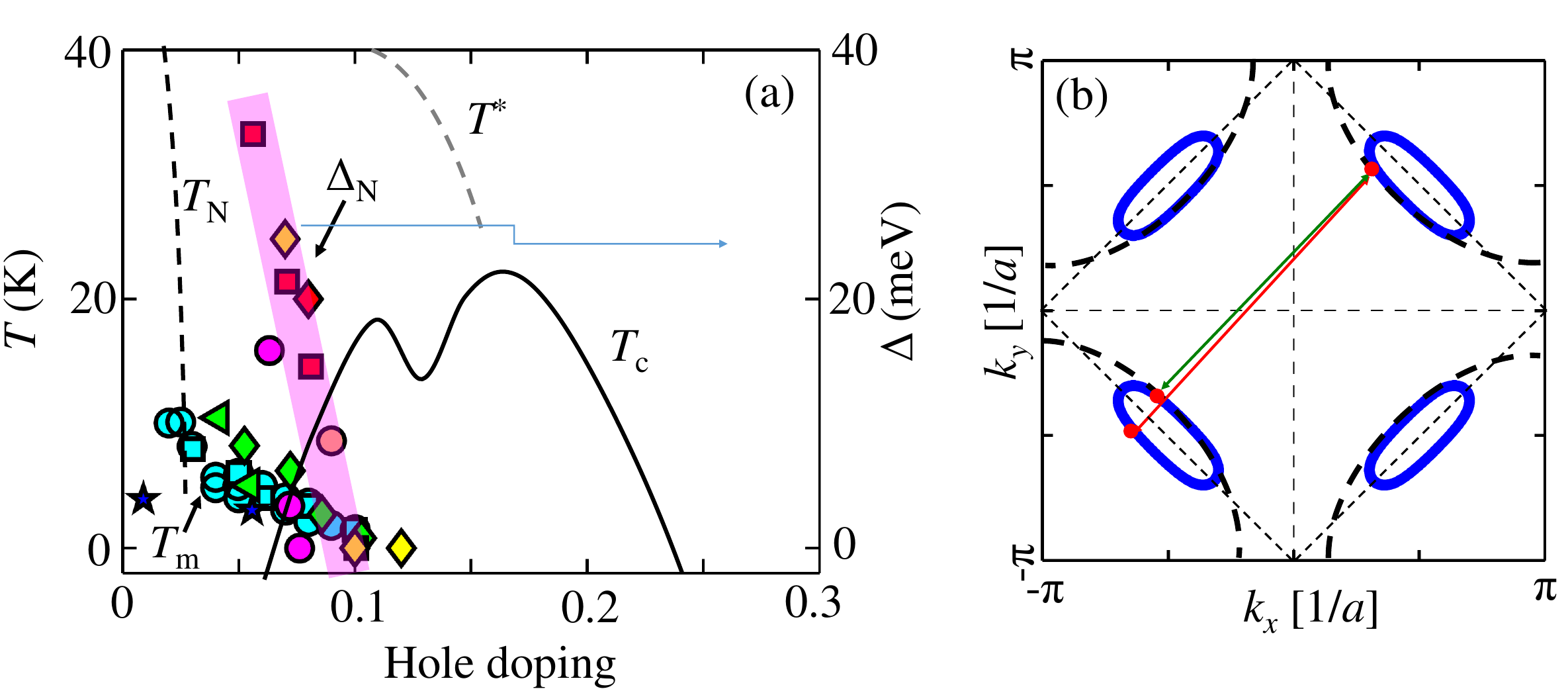}}
\caption{(a) A phase diagram of hole-doped cuprates, with various phases and gaps. Blue, green, and cyan symbols give the spin-glass temperature,\cite{Tsg} as well as the (short-range) SDW temperature (collectively denoted by $T_m$)\cite{NMRBi11,NMRBi12,NMRTlHg12,NMRYBCO12,NeutronYBCO10,NeutronYBCO11,twodomes}  in various hole-doped cuprates. Red, magenta, and yellow symbols are the nodal gap for different systems\cite{LSCO2000,ccoc,bi22122006,bi2212,LSCO2013,bi2201,YBCO}. $T_{\rm N}$, $T_{\rm c}$, and $T^*$ are N\'eel, SC transition, and the pseudogap temperatures, respectively. Further description of each data is given in Ref.~\cite{twodomes}. Notably, both $T_{\rm m}$ and the nodal gap $(\Delta_{\rm N})$ disappear around the same doping ($\sim$10\%) in these materials, suggesting a coexistence of them.  (b) Fermi surface topology of the underdoped cuprates without (black dashed line), and with the AFM order (solid blue line) at $x\sim$0.08. Thin gray line draws the reduced BZ for the wavevector ${\bf Q}=(\pi,\pi)$ which is half of the paramagnetic BZ. Double green (red) arrows depict the $d$-wave (PWD) SC channel. 
}
\label{fig1}
\end{figure}

Our theory relies on the fact that the N\'eel order in the underdoped cuprates smoothly transforms into the spin-glass or short-ranged spin-density-wave (SDW) state which extents upto 10\% doping where nodeless SC gap is also observed. The evidence of the spin-glass\cite{Tsg} and SDW state\cite{twodomes,NMRBi11,NMRBi12,NMRTlHg12,NMRYBCO12,NeutronYBCO10,NeutronYBCO11}  is presented in various experimental probes and also in various hole doped cuprates. Optical studies also showed that a gap persists in the far-infrared (FIR) region.\cite{FIRgap} We denote these gaps collectively by $T_{\rm m}$ in Fig.~\ref{fig1}(a), which clearly extends to the SC region. Interestingly, the nodal gap observed in various ARPES data in different materials also extends up to $x\sim $0.1 and disappears with $T_{\rm m}$. The interplay between a SC order with a translational symmetry breaking competing order generically yields a dynamically generated new SC order parameter with finite center-of-mass momentum. Earlier calculations predicted that for the competition of AFM and $d$-wave SC, the third SC order parameter is a nodal $d$-wave superconductor if no magnetic field is applied,\cite{dSCtriplet,so5} or a time-reversal symmetry breaking $\pi$ triplet SC if a magnetic field is applied in the heavy-fermion systems\cite{cecoin5}. 

The $d$-wave singlet SC and the staggered AFM order parameters are defined as $\langle c^{\dag}_{{\bf k}\uparrow} c^{\dag}_{-{\bf k}\downarrow}\rangle$, and $\langle c^{\dag}_{{\bf k}+{\bf Q}\uparrow} c_{{\bf k}\uparrow}-c^{\dag}_{{\bf k}+{\bf Q}\downarrow} c_{{\bf k}\downarrow}\rangle$, where $c^{\dag}_{{\bf k}\uparrow}$ is the fermionic creation operator at the crystal momentum ${\bf k}$ with up spin (say). Having these two order parameters at hand, the third order parameter must be SC as defined by $\langle c^{\dag}_{{\bf k}\uparrow} c^{\dag}_{-{\bf k}-{\bf Q}\downarrow}\rangle$. This is a finite momentum pair with its center-of-mass momentum determined by ${\bf Q}$. Let us assume the corresponding structure factor for this PDW pair is $g_{\bf k}$. The interesting part is that as we interchange two fermions in this PDW pair, the structure factor changes to $g_{{-\bf k}-{\bf Q}}$. Given that the pairing stems from unconventional mechanism, anisotropic and sign-reversal pairing channel is anticipated. Given that ${\bf Q}=(\pi,\pi)$, for even parity channel (such as $d$-wave), we have $g_{{-\bf k}-{\bf Q}}=-g_{\bf k}$, requiring a spin-triplet  pairing (not compatible). On the other hand, an odd-parity structure factor such as $p+ip$ - wave state gives $g_{{-\bf k}-{\bf Q}}=g_{-\bf k}$ with a spin-singlet channel. We remind that for a zero-momentum pairing, $p+ip$ state is a odd-parity, and spin-triplet pairing. In this way the present case is a unique counter example to the existing odd parity pairing literature. Such a $p+ip$ pairing channel also fits well with the observation of the fully gapped superconductivity in these compounds at small dopings. 

We use a mean-field Hamiltonian based on the short range Coulomb repulsion and attractive potential to self-consistently evaluate all three coupled order parameters at a given doping. For experimental verification of our theory, we compute the temperature dependence of the superfluid density (inversely proportional to the square of the magnetic penetration depth) in the uniform coexistence state of all three interactions. We show that the exponential temperature dependence of the superfluid density will be a definitive test of the SC origin of the fully gapped state in lightly doped cuprates. Furthermore, the amplitude modulation of the second SC gap can be directly visualized by scanning tunneling microscopy or spectroscopy (STM/STS).

The rest of the paper is arranged as follows. In Sec.~II, we discuss the mean-field theory for the three competing orders and their symmetry properties. In Sec.~III, we present our results for the evolution of the fully gapped electronic structure in the PDW state and the superfluid density calculation to show the exponential behavior in the low-temperature region in this state. Finally, we conclude in Sec.~IV. Further details of the mean-field theory, derivation of self-consistent order parameters and the superfluid calculations are given in Appendix. 

\section{Theory}

Our starting Hamiltonian is a single band Hubbard model with two pairing interactions $V_{1,2}$: 
\begin{eqnarray}\label{tHam}
H&=&\sum_{{\bf k},\sigma}\xi_{{\bf k}}c^{\dag}_{{\bf k}\sigma}c_{{\bf k}\sigma}
+U\sum_{{\bf k},{\bf k}^{\prime}}
c^{\dag}_{{\bf k}\uparrow}c_{{\bf k}+{\bf Q}\uparrow}
c^{\dag}_{{\bf k}^{\prime}-{\bf Q}\downarrow}c_{{\bf k}^{\prime}\downarrow}\nonumber\\
&&+\sum_{{\bf k},{\bf k}^{\prime}}V_{1,{\bf k}{\bf k}^{\prime}}
c^{\dag}_{{\bf k}\uparrow}c^{\dag}_{-{\bf k}\downarrow}c_{-{\bf k}^{\prime}\downarrow}
c_{{\bf k}^{\prime}\uparrow}\nonumber\\
&&+\sum_{{\bf k}{\bf k}^{\prime}}V_{2,{\bf k}{\bf k}^{\prime}}
c^{\dag}_{{\bf k}\uparrow}c^{\dag}_{-{\bf k}-{\bf Q}\downarrow}c_{-{\bf k}^{\prime}-{\bf Q}\downarrow}
c_{{\bf k}^{\prime}\uparrow}.\nonumber\\
\end{eqnarray}
Here $\xi_{\bf k}$ is the bare dispersion, $U$ is the onsite Coulomb interaction, $V_{1,2}$ are the pairing strengths for $d$-wave and $p$-wave superconductivity, respectively. We model the bare dispersion in the tight-binding form as $\xi_{{\bf k}}=-2t(\cos{k_x}+\cos{k_y})-4t^{\prime}\cos{k_x}\cos{k_y}-2t^{\prime\prime}(\cos{2k_x}+\cos{2k_y})-4t^{\prime\prime\prime}(\cos{2k_x}\cos{k_y}+\cos{k_x}\cos{2k_y})-E_F$, where $(t,~t^{\prime},~t^{\prime\prime},~t^{\prime\prime\prime},~E_F)$=(250,-25, 12, 35, -155) in meV which is obtained by fitting to the dispersion obtained by ARPES in LSCO.\cite{BobTB} $c_{{\bf k},\sigma}$ is the fermion annihilation operator at crystal momentum ${\bf k}$ with spin $\sigma=\pm$. ${\bf Q}=(\pi,\pi)$ is the AFM nesting vector in 2D. We assume here a commensurate AFM order so that ${\bf k}+{\bf Q}={\bf k}-{\bf Q}$. The staggered spin magnetization is defined as $m=\sum_{{\bf k},\sigma}\sigma\Big\langle c^{\dag}_{{\bf k}+{\bf Q}\sigma}c_{{\bf k}\sigma}\Big\rangle$. According to the SO(5) symmetry between the AFM and $d$-wave superconductivity,\cite{so5} the Lie Algebra of the group generators, that is, the order parameters, dictate that the commutator of any two order parameters generate a third one. Using such a symmetry, the spatial and spin symmetry of each parameter can be easily deduced (see below). As discussed above, if $V_1$ is assumed to be singlet $d$-wave symmetry, the AFM state guarantees that $V_2$ should be spin-singlet $p+ip$-wave pairing. The singlet interactions are $V_{1,{\bf k}{\bf k}^{\prime}}=V_{10}d_{{\bf k}}d_{{\bf k}^{\prime}}$ and $V_{2,{\bf k}{\bf k}^{\prime}}=V_{20}p_{{\bf k}}p_{{\bf k}^{\prime}}$, where $d_{{\bf k}}=\cos{k_x}-\cos{k_y}$ and $p_{{\bf k}}=\sin{k_x}+i\sin{k_y}$. We assume that $V_{10,20}$ are attractive. Then the $d$-wave SC gap is defined as 
\begin{equation}
\Delta_{1,{\bf k}}=\Delta_{10}d_{{\bf k}}=V_{10}d_{{\bf k}}\sum_{{\bf k}^{\prime}}d_{{\bf k}^{\prime}}\Big\langle c^{\dag}_{{\bf k}^{\prime}\uparrow}c^{\dag}_{-{\bf k}^{\prime}\downarrow}\Big\rangle.
\end{equation}
For the $d$-wave, $\Delta_{1{\bf k}+{\bf Q}}-\Delta_{1,{\bf k}}$. The PDW gap is defined as
\begin{equation}
\Delta_{2,{\bf k}}=\Delta_{20}p_{{\bf k}}=V_{20} p_{{\bf k}}\sum_{{\bf k}^{\prime}}p_{{\bf k}^{\prime}}\Big\langle c^{\dag}_{{\bf k}+{\bf Q}\uparrow}c^{\dag}_{-{\bf k}\downarrow}\Big\rangle.
\end{equation}
Similarly, for $p+ip$-wave we have $\Delta_{2,{\bf k}+{\bf Q}}=-\Delta_{2,{\bf k}}$. The $p+ip$ pairing is time-reversal invariant. To see that we write down the Ginzburg-Landau type Free energy as:
\begin{eqnarray}\label{GL}
F = \alpha \left[ \Delta_{10} ({\bf \Delta}_{20}^{*}\cdot{\bf m}) + \Delta_{10}^{*} ({\bf \Delta}_{20}\cdot{\bf m})\right] + ...,
\end{eqnarray}
where $\alpha$ is a constant and `...' represents all the other quadratic, quartic and dipole coupling terms. Generally, three components of the $O(3)$ magnetic moment `${\bf m}$' correspondingly can generate three distinct PDW states, denoted by a ${\bf \Delta}_{20}$ vector, even for a single component zero-momentum pairing gap $\Delta_{10}$. In the present case, we only have a SDW state for $m_z$, and hence a single component of $\Delta_{20}$. Furthermore, as mentioned before, since the time-reversed component of $\Delta_{20}$ lies outside the magnetic zone boundary with an opposite sign, the above coupling remains invariant under time-reversal symmetry. Finally, since superconductivity breaks gauge symmetry, while magnetism does not, the third order parameter must also break the gauge symmetry, i.e., the third order parameter must be a SC phase which allows the Hamiltonian to be gauge invariant. 

We express the mean-field Hamiltonian in the Nambu-Gor'kov basis in the magnetic BZ as $\Psi_{{\bf k}}=(c_{{\bf k}\uparrow},~c_{{\bf k}+{\bf Q}\uparrow},~c^{\dag}_{-{\bf k}\downarrow},~c^{\dag}_{-{\bf k}-{\bf Q}\downarrow})^T$:
\begin{eqnarray}\label{Ham}
\Big\langle \Psi_{{\bf k}}^{\dag}|H|\Psi_{{\bf k}}\Big\rangle=
\begin{centering}\left(\begin{array}{cccc}
\xi_{{\bf k}} & -Um & \Delta_{1{\bf k}} &\Delta_{2{\bf k}}\\
-Um & \xi_{{\bf k}+{\bf Q}} & -\Delta_{2{\bf k}} & -\Delta_{1{\bf k}}\\
\Delta^*_{1{\bf k}} & -\Delta^*_{2{\bf k}} &-\xi_{{\bf k}} &-Um\\
\Delta^{*}_{2{\bf k}} &-\Delta^*_{1{\bf k}}&-Um&-\xi_{{\bf k}+{{\bf Q}}}
\end{array}\right).\end{centering}
\end{eqnarray}
We diagonalize the Hamiltonian in three steps by three consecutive Bogolyubov transformations. First we diagonalize the PDW SC gap with the change of basis as $c_{{\bf k}\uparrow}=f_{{\bf k}}t_{{\bf k}\uparrow}+g_{{\bf k}}t^{\dag}_{-{\bf k}\downarrow}$, $c^{\dag}_{-{\bf k}-{\bf Q}\downarrow}=g^*_{{\bf k}}t_{{\bf k}\uparrow}+f^*_{{\bf k}}t^{\dag}_{-{\bf k}\downarrow}$, where $t_{{\bf k}\sigma}$ are the Bogolyubov operators, and the corresponding coherence factors are
$2|f_{{\bf k}}|^2=\left(1+\frac{\xi_{{\bf k}}^+}{E_{{\bf k}}^{\pi}}\right)$, and $2|g_{{\bf k}}|^2=\left(1-\frac{\xi_{{\bf k}}^+}{E_{{\bf k}}^{\pi}}\right)$. The corresponding dispersion energies are $E_{{\bf k}}^{\pi}=\pm\sqrt{(\xi_{{\bf k}}^+)^2+|\Delta_{2{\bf k}}|^2}$, with $\xi_{{\bf k}}^{\pm}=(\xi_{{\bf k}}\pm\xi_{{\bf k}+{\bf Q}})/2$. In this new basis, the effective complex AFM gap and $d$-wave SC gap can be expressed by $\Delta_{{\bf k}}^m=(Um\xi_{{\bf k}}^+ +\Delta_{2{\bf k}}\Delta_{1{\bf k}})/E_{{\bf k}}^{\pi}$ , and $\Delta_{{\bf k}}^{\rm SC}=(-\Delta_{1{\bf k}}\xi_{{\bf k}}^+ + \Delta_{2{\bf k}}Um)/E_{{\bf k}}^{\pi}$. With the effective gaps, the above Hamiltonian reduces to a typical AFM+$d$-SC phase which can now be diagonalized in the same fashion as done previously in Refs.~\cite{TanmoyED,Tanmoytwogap} (see Appendix). The corresponding effective AFM and SC coherence factors, respectively, are $2|\alpha_{{\bf k}}|^2=\left(1+\frac{\xi_{{\bf k}}^-}{E_{{\bf k}}^m}\right)$, $2|\beta_{{\bf k}}|^2=\left(1-\frac{\xi_{{\bf k}}^-}{E_{{\bf k}}^m}\right)$, $2|u^{\pm}_{{\bf k}}|^2=\left(1+\frac{E_{{\bf k}}^{\pi}\pm E_{{\bf k}}^{m}}{E_{{\bf k}}^{\pm}}\right)$, and $2|v^{\pm}_{{\bf k}}|^2=\left(1-\frac{E_{{\bf k}}^{\pi}\pm E_{{\bf k}}^{m}}{E_{{\bf k}}^{\pm}}\right)$. The quasiparticle energies are $(E_{{\bf k}}^{\pm})^2=(E_{{\bf k}}^{\pi}\pm E_{{\bf k}}^{m})^2 + |\Delta^{\rm SC}_{{\bf k}}|^2$, where $E_{{\bf k}}^m=\sqrt{(\xi_{{\bf k}}^-)^2 +|\Delta^m_{{\bf k}}|^2}$. Clearly, in the present case, the eigenvalues are particle-hole symmetric. 

We evaluate all three order parameters and the chemical potential self-consistently for given values of constant interactions $U$, $V_{10}$ and $V_{20}$ at a given doping by solving the following coupled equations:
%
\begin{eqnarray}\label{orderparam}
m&=&\sum_{{\bf k}}^{\prime}\alpha_{{\bf k}}\beta_{{\bf k}}(|f_{\bf k}|^2-|g_{\bf k}|^2)
\Big[((v_{{\bf k}}^{-})^2-(v_{{\bf k}}^{-})^2)\nonumber\\
&&\times n(E_{{\bf k}}^{+})-((v_{{\bf k}}^{-})^2-(u_{{\bf k}}^{-})^2)n(E_{{\bf k}}^{-})\Big],\\
\Delta_{10}&=&V_{10}\sum^{\prime}_{{\bf k}}\sum_{\nu=\pm}d_{{\bf k}}\Big[{\rm Re}[f^2_{{\bf k}}-g^2_{{\bf k}}]u_{{\bf k}}^{\nu}v_{{\bf k}}^{\nu}\nonumber\\
&&+2{\rm Re}[f_{{\bf k}}g_{{\bf k}}]\alpha_{{\bf k}}\beta_{{\bf k}}((u^{\nu}_{{\bf k}})^2-(v^{\nu}_{{\bf k}})^2)\Big]\tanh{(\beta E_{{\bf k}}^{{\nu}}/2)},\\
\Delta_{20}&=&V_{20}\sum^{\prime}_{{\bf k}}p_{{\bf k}}\sum_{\nu=\pm}\Big[\big[{\rm Re}[f_{{\bf k}}g_{{\bf k}}]((u^{\nu}_{{\bf k}})^2-(v^{\nu}_{{\bf k}})^2)\nonumber\\
&&+2\alpha_{{\bf k}}\beta_{{\bf k}}u_{{\bf k}}^{\nu}v_{{\bf k}}^{\nu}(|f_{{\bf k}}|^2-|g_{{\bf k}}|^2)\big]\tanh{(\beta E_{{\bf k}}^{\nu}/2)})\Big].
\end{eqnarray}
%
Here the prime over a summation indicates that the summation is restricted within the magnetic BZ. $n(E^{\nu}_{{\bf k}})$ is the Fermi-Dirac function.

For $U=3t$ ($t$ is the nearest neighbor tight-binding hopping parameter), $V_{10/20}$ = -87, -60 meV, we get $m$ = 0.1, $\Delta_{10/20}$ = -20, -10 meV, respectively at doping $x$ = 0.08 in LSCO which are realistic gap values for this material.\cite{LSCO2013} The resulting FS in the AFM state without superconductivity is shown in Fig.~\ref{fig1}(b). However, to facilitate the visualization of the nature of gap openings in the electronic structure, we use artificially large value of $\Delta_{10}$ = 100~meV and $\Delta_{20}$ = 50~meV in Fig.~\ref{fig2}(a). In the presence of three competing interactions, the CuO$_2$ antibonding band splits into four quasiparticle states with finite gap at everywhere in the BZ. The bands are colored with corresponding filling factor which adds to 1 at each momentum. At the nodal point, the band gap is purely determined by the PDW pairing, since a AFM + $d$-wave pairing gives nodal states there. And the gap at the antinodal region is a mixture of all three gaps determined by $\sqrt{|\Delta_{{\bf k}}^{\rm m}|^2+|\Delta_{{\bf k}}^{\rm SC}|^2}$.

To experimentally verify the SC origin of the fully gapped state, we calculate the penetration depth ($\lambda(T)$) or superfluid density via linear response theory by accounting for both the diamagnetic current (proportional to number of SC quasiparticles) and residual paramagnetic current (carried by normal state electrons) due to external magnetic field. The detailed derivation is given in the Appendix, and we here write down the final expression:
%
\begin{eqnarray}\label{lambda1}
&&\lambda_{ij}^{-2}(T)=\frac{4\pi e^2}{c^2\Omega}\nonumber\\
&&\times \sum_{{\bf k},\nu=\pm}\left[\left(\frac{1}{M^{\nu}_{{\bf k}i,j}}\right)
\left(1-\frac{E^{\pi}_{{\bf k}}+E^{m}_{{\bf k}}}{E^{\nu}_{{\bf k}}}\tanh{(\beta E^{\nu}_{{\bf k}}/2)}\right)\right. \nonumber\\
&&~~~ +\left(\frac{1}{m^{\nu}_{{\bf k}i,j}}\right)\tanh{(\beta E^{\nu}_{{\bf k}}/2)}
\left. -\frac{\beta}{2}V^{\nu}_{{\bf ki}}V^{\nu}_{{\bf kj}}{\rm sech}^2(\beta E_{{\bf k}}^{\nu}/2)\right].\nonumber\\
\end{eqnarray}
%
Here $e$, $c$ are bare electronic charge and speed of light respectively, and $\Omega$ is the unit cell volume. $M$, $m$, and $V$ are different band masses, and velocity defined in Appendix (note that we use the same notation of $m$ for magnetization and band mass, and the context of it's appearance clarifies its meaning). Due to $C_4$ symmetry in the system, the cross-terms are zero and the penetration depth matrix is diagonal. Therefore, we plot $\lambda_{xx}^{-2}(T)$ component only in Fig.~\ref{fig4} and normalized to its zero-$T$ value.

\section{Results}

\begin{figure}[t]
\hspace{-0.5cm}
\rotatebox[origin=c]{0}{\includegraphics[width=.99\columnwidth]{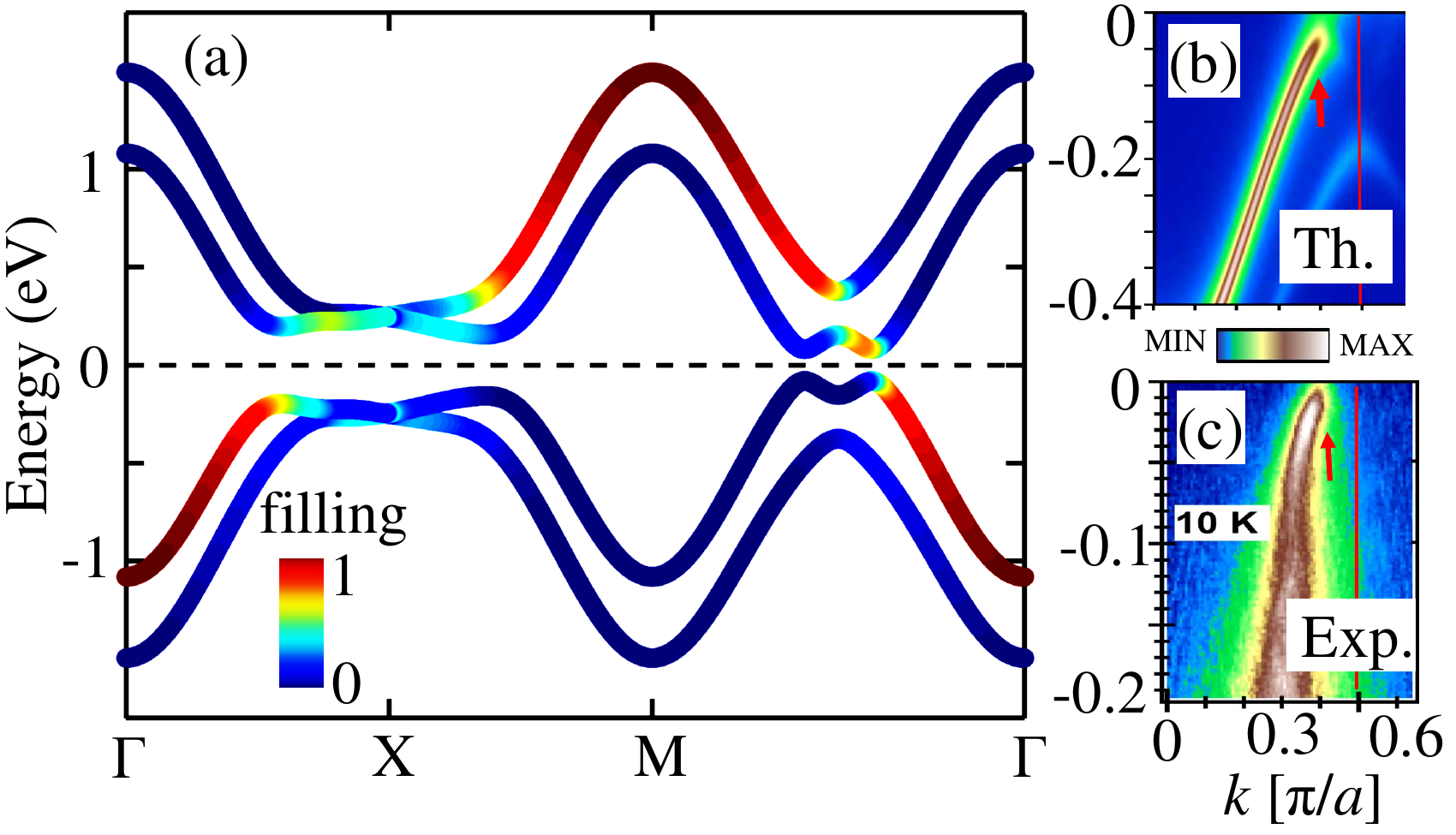}}
\caption{(Color online) (a) Computed electronic dispersion in the coexistence state of AFM, $d$-wave SC and PDW type $p+ip$-wave SC. 
(b) Spectral weight map of the same spectrum along the nodal direction, exhibiting the location of band bending far below the magnetic BZ (arrows). The red vertical line marks the magnetic BZ boundary along the nodal axis. This result is consistent with the experimental observation of the same system, reproduced in (c) from Ref.~\onlinecite{LSCO2013}. 
}
\label{fig2}
\end{figure}

A spectroscopic method to differentiate between the SC and magnetic origins of the nodal gap is to observe whether the gapped quasiparicle state has a band-folding or band-bending at the magnetic zone boundary or at the normal state Fermi momentum. To directly compare our result with the angle-resolved photoemission spectroscopy (ARPES) data, we compute the single-particle spectral weight along the nodal direction as shown in Fig.~\ref{fig2}(b), while Fig.~\ref{fig2}(c) is the ARPES data\cite{LSCO2013} for the same sample. In both cases, we see that the band folding (indicated by vertical arrows) occurs at a momentum which is far below the magnetic BZ boundary at ${\bf k}=(\pi/2,\pi/2)$ point (red line). The additional band appeared in the theory is a shadow band induced by the SC order, and possess much weak intensity when the SC gap is reduced to the realistic value of $\Delta_{20} \sim$ 10~meV.

\begin{figure}[t]
\rotatebox[origin=c]{0}{\includegraphics[width=.99\columnwidth]{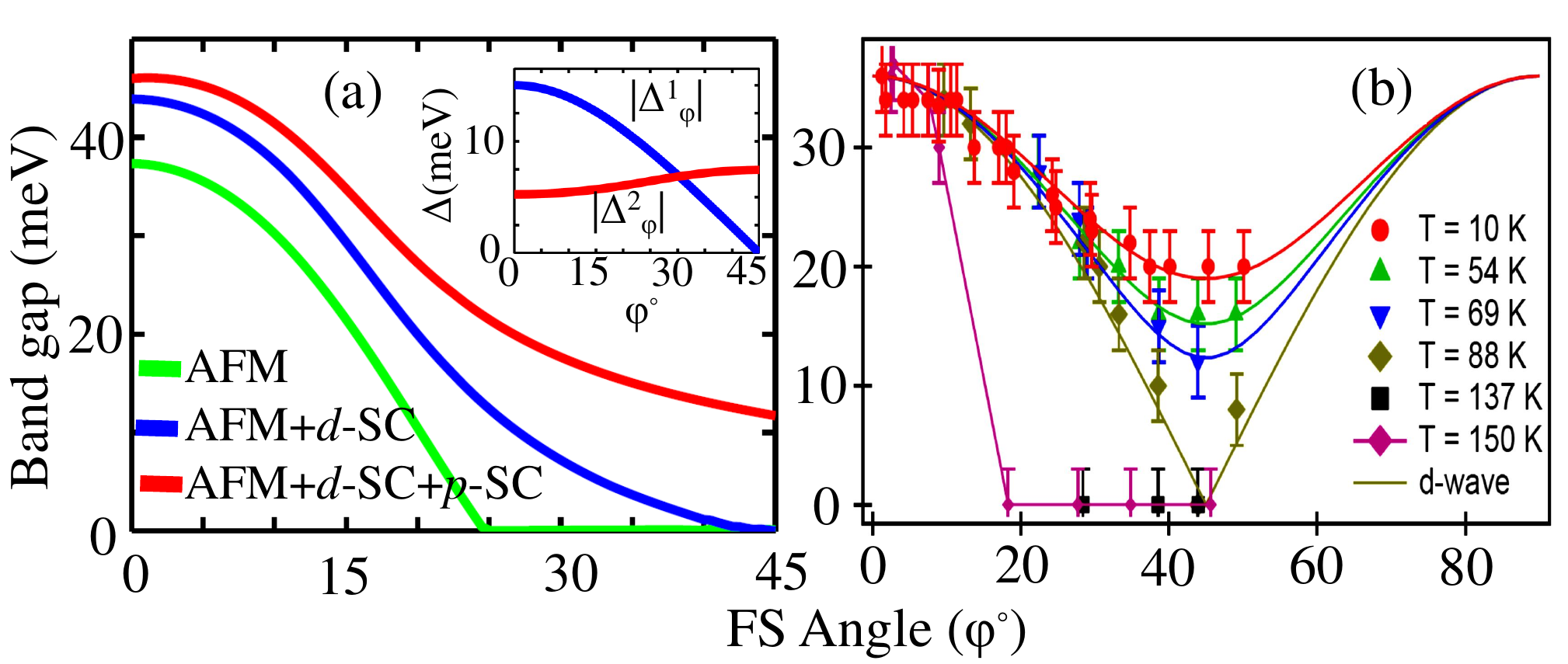}}
\caption{(Color online) (a) Extracted band gap (not the order parameter) as a function of the FS angle $\phi^o$, where $\phi=0^o$ and $\phi=45^o$ give antinodal and nodal directions, respectively. 
(b) The corresponding experimental result, reproduced from Ref.~\onlinecite{LSCO2013} for the same sample, are in detailed agreement with the theory.}
\label{fig3}
\end{figure}

In Fig.~\ref{fig3}(a), we plot the extracted band gap (not the individual order parameters) as a function of FS angle ($\phi^o$) measured with respect to the $k_x=\pi$ axis centering at M$=(\pi,\pi)$ point. We see that in the non-SC, and AFM metallic state ($\Delta_{10,20}=0$, but magnetization $m$ = finite), gapless Fermi pocket exists in the region of $\phi\sim 22^o-45^o$. The corresponding hole-pocket FS is also seen in the ARPES data for the same sample above the SC transition temperature $T_c$ [reproduced from Ref.~\onlinecite{LSCO2013} in Fig.~\ref{fig3}(b)]. In the case of a pure $d$-wave SC without any PDW component ($\Delta_{20}=0$ and the other two order parameters are finite), the electronic state is gapped throughout the BZ except at $\phi=45^o$. The non-monotonic nature of the band gap is reproduced in such a case, and has been observed in ARPES measurements (see, e.g., Ref.~\onlinecite{Tanmoytwogap}). However, when the PDW component is turned on, a finite gap opens everywhere as described by $\sqrt{|\Delta_{{\bf k}}^{\rm m}|^2+|\Delta_{{\bf k}}^{\rm SC}|^2}$. It is interesting to notice that the odd-parity spin-singlet PDW SC of the present form is fully gapped throughout the BZ [see {\it inset} to Fig.~\ref{fig3}(a)], regardless of the topology of the underlying FS and is argued to be topologically non-trivial if time-reversal symmetry is retained,\cite{LFu} which is also applicable in the present case. The obtained gap structure agrees remarkably well with that observed in the ARPES measurement in the same sample as shown in Fig.~\ref{fig3}(b) at low temperatures.

\begin{figure}[t]
\rotatebox[origin=c]{0}{\includegraphics[width=.9\columnwidth]{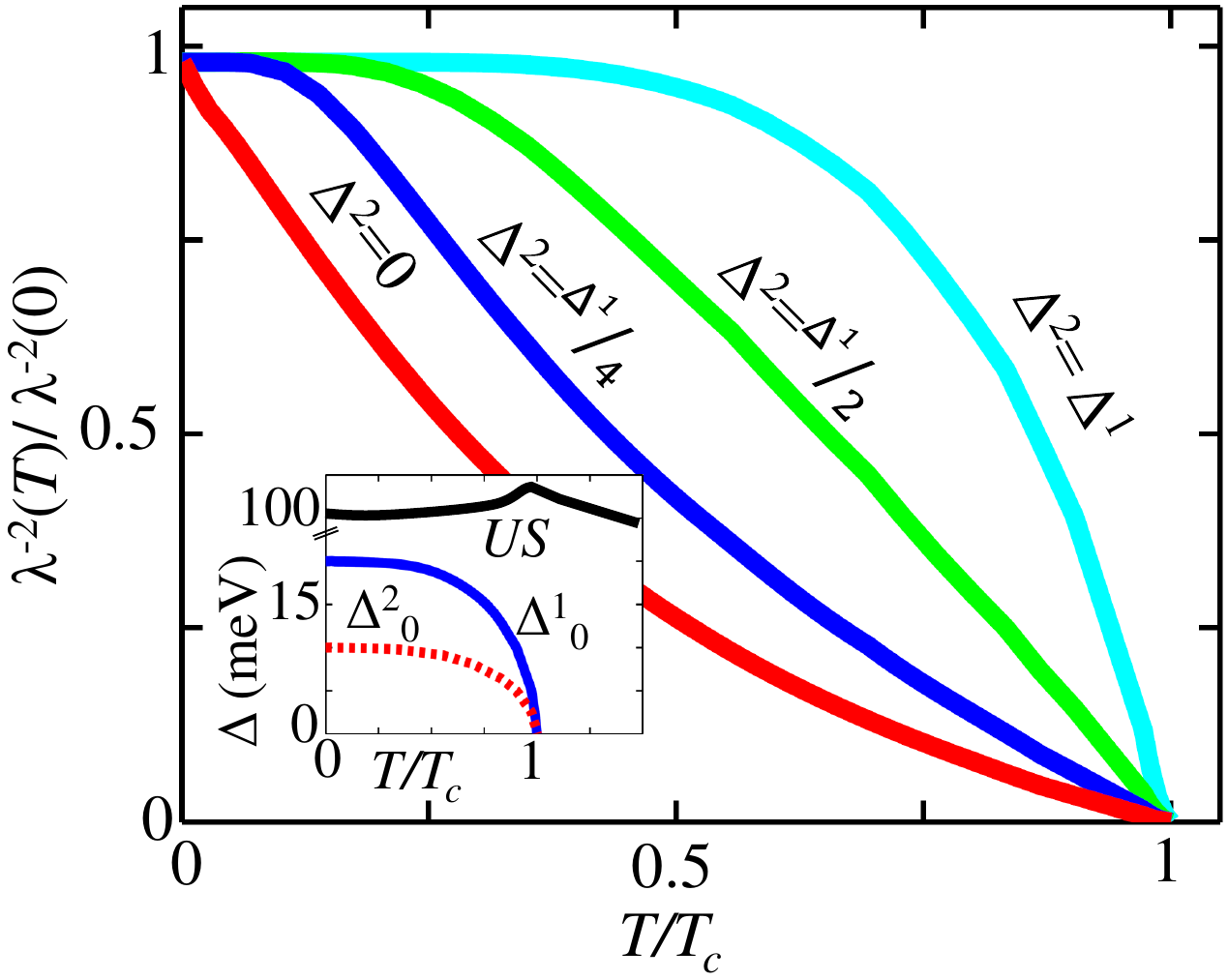}}
\caption{(Color online) Computed $\lambda^{-1}(T)$
for several representative cases as indicated by adjacent labeling to each plot. 
{\it Inset:} Self-consistently calculated AFM gap, $d$- and $p$-wave gaps.}
\label{fig4}
\end{figure}

Finally, we compute the penetration depth for different cases of the competition of the phases under study to verify their existences in the underlying spectrum by using Eq.~\eqref{lambda1}. It is well known that for a fully gapped superconductor, the inverse squared penetration depth $\lambda^{-2}(T)$ (proportional to the superfluid density) shows exponential $T$ - dependence in the low - $T$ region, where nodal $d$ - wave SC exhibits a linear-in-$T$ behavior, which sometimes modifies to the $T^2$ - dependence due to disorder. 

To get a quantitative understanding, we calculate the temperature dependence of all gap values self-consistently at fixed values of the interaction parameters and at a fixed doping. Expectedly, we see that for a pure nodal $d$-wave ($\Delta_{20}=0$), the linear-in-$T$ dependence of $\lambda^{-2}$ is evident even in the presence of AFM ground state. With slowly turning on the fully gapped PDW pairing, the exponential dependence of $\lambda^{-2}$ appears whose $T$-range of flat or exponential region depends on the relative strength of $\Delta_{20}/\Delta_{10}$. For $\Delta_{10}=\Delta_{20}$, the computed penetration depth shows a fully exponential behavior at all temperatures, which is a hallmark feature of the nodeless SC gap. The transition from a pure $d$-wave like $\lambda_{ij}^{-2}(T)$ behavior to nodeless $d$-wave behavior is reported in theory and experimental data of electron doped cuprates.\cite{TanmoyED} Since exponential behavior only appears in the very low-temperature (10-15\% of $T_c$), and the fully gapped structure survives only in the underdoped region where $T_c$ is low, very high-resolution experimental facility is required detect such a behavior. Therefore, Bi2212, having higher $T_c$ among others where fully gapped SC state is observed, will be an ideal system to experimentally carry out the penetration depth measurement for this purpose.

\section{Discussion and conclusion}

The present paper concerns with a mean-field and self-consistent treatment of three order parameters. A matter of immediate concern can be to what extend the mean-field theory is valid in the underdoped cuprates. It is also noted in Fig.~1b, that the AFM order becomes short-ranged in reaching to the SC dome, while the theory continues to assume long-ranged order. Typically, the correlation length of the long-range magnetic order can be reduced due to quantum fluctuations and/or disorder, among other effects. On these two accounts, we argue that both the effects are experimentally known to be substantially less on the AFM order. Firstly, the enhanced quantum fluctuations in reaching the quantum critical point typically manifests into a non-Fermi liquid behavior as exhibited in a linear resistivity-temperature exponent. On the other hand, in cuprates the resistivity-temperature exponent is close to 2 near the AFM critical point\cite{twodomes} $-$ demonstrating that the quasiparticle description within the Fermi liquid mean-field model is sufficient to describe the low energy features. Secondly, Alloul {\it et al.}\cite{Alloul} showed that with sufficient disorder substitution in cuprates, the the SC dome splits into two domes, while the AFM temperature remains considerably immune to the disorder effect. Therefore, the effects of disorder can also be ignored in the calculation aiming at proving the salient features arising from the competition of the AFM and superconductivity in the underdoped cuprates. 

In conclusion, the present paper provides a generalized framework for the phase stability of the AFM and $d$-wave superconductivity in hole-doped cuprates. Following the SO(5) theory,\cite{so5} we show that the uniform coexistence of these two phases is accompanied by a PDW order parameter. The symmetry of the ${\bf Q}=(\pi,\pi)$ value AFM state and $d$-wave pairing restricts the symmetry of PDW state to be spin singlet, odd parity gap function which commences a fully gapped quasiparticle spectrum. The odd parity order parameter breaks spin-rotational symmetry, but via coupling to the AFM state, it respects time-reversal symmetry. Interestingly, such a fully gapped odd parity pairing symmetry in a time-reversal invariant condition is called topological superconductor.\cite{LFu} In general, such a PDW term can be expected to be negligibly small, however, the critical fluctuations of the parent phases can significantly enhance the strength of the PDW pairing interaction. While finite-momentum SC pairing has been proposed in many Pauli limited superconductors, such as nodal $d$-wave in CeCoIn$_5$ at high magnetic field,\cite{cecoin5} and $p$-wave in the Rashba-type spin-orbit coupling systems without any realization to date, underdoped cuprates provide a clean and zero field platform to discover this exotic phase, if exists.

\begin{acknowledgments}
We thank C. Panagopolous for suggesting several references for the spin glass and spin density wave state in underdoped cuprates. The work is supported by the SERC computer facility and the IISc startup grant.
\end{acknowledgments}

\appendix

\section{Derivation of the Hamiltonian}
From the study of the group algebra\cite{so5,so8}, it is obvious that the coexistence of the any two non-commutating order parameter produce a third order parameter. In case of AFM and $d$-superconductivity, there should be a third, dynamically generated, order parameter.\cite{kyung} The strength of this third order parameter is subjects to the corresponding coupling constant. In the present work, we develop the theory for how the coexistence of AFM and $d$-wave superconductivity can generate a non-zero center of mass SC order parameter. After substituting these mean-field orders, the total Hamiltonian reads,
\begin{widetext}
\begin{eqnarray}\label{tHam}
H&=&\sum_{{\bf k}\sigma}\xi_{{\bf k}}c^{\dag}_{{\bf k}\sigma}c_{{\bf k}\sigma}
-Um\sum_{{\bf k},\sigma}\sigma
c^{\dag}_{{\bf k}+{\bf Q}\sigma}c_{{\bf k}\sigma}\nonumber\\
&&+\sum_{{\bf k}}\Delta_{1{\bf k}}\Big(
c^{\dag}_{{\bf k}\uparrow}c^{\dag}_{-{\bf k}\downarrow}+c_{-{\bf k}\downarrow}
c_{{\bf k}\uparrow}\Big)
+\sum_{{\bf k}}\Delta_{2{\bf k}}\Big(
c^{\dag}_{{\bf k}\uparrow}c^{\dag}_{-{\bf k}-{\bf Q}\downarrow}+c_{-{\bf k}-{\bf Q}\downarrow}
c_{{\bf k}\uparrow}\Big)\nonumber\\
%
&=&\sum^{\prime}_{{\bf k},\sigma}\xi_{{\bf k}}^+\Big(c^{\dag}_{{\bf k}\sigma}c_{{\bf k}\sigma}
+c^{\dag}_{{\bf k}+{\bf Q}\sigma}c_{{\bf k}+{\bf Q}\sigma}\Big)
+\xi_{{\bf k}}^-\Big(c^{\dag}_{{\bf k}\sigma}c_{{\bf k}\sigma}
-c^{\dag}_{{\bf k}+{\bf Q}\sigma}c_{{\bf k}+{\bf Q}\sigma}\Big)
-2Um\sum^{\prime}_{{\bf k},\sigma}\sigma
c^{\dag}_{{\bf k}+{\bf Q}\sigma}c_{{\bf k}\sigma}\nonumber\\
&&+\sum^{\prime}_{{\bf k}}\Delta_{1{\bf k}}\Big(
c^{\dag}_{{\bf k}\uparrow}c^{\dag}_{-{\bf k}\downarrow}+c_{-{\bf k}\downarrow}
c_{{\bf k}\uparrow}-c^{\dag}_{{\bf k}+{\bf Q}\uparrow}c^{\dag}_{-{\bf k}-{\bf Q}\downarrow}
-c_{-{\bf k}-{\bf Q}\downarrow}
c_{{\bf k}+{\bf Q}\uparrow}\Big)\nonumber\\
&&+\sum^{\prime}_{{\bf k}}\Delta_{2{\bf k}}\Big(
c^{\dag}_{{\bf k}\uparrow}c^{\dag}_{-{\bf k}-{\bf Q}\downarrow}+c_{-{\bf k}-{\bf Q}\downarrow}
c_{{\bf k}\uparrow}
-c^{\dag}_{{\bf k}+{\bf Q}\uparrow}c^{\dag}_{-{\bf k}\downarrow}-c_{-{\bf k}\downarrow}
c_{{\bf k}+{\bf Q}\uparrow}\Big) 
+ Um^2 + \frac{|\Delta_{10}|^2}{V_{10}}+ \frac{|\Delta_{20}|^2}{V_{20}}.
\end{eqnarray}
\end{widetext}
The prime over the summation indicate that the sum is defined in the magnetic BZ. Defining a Nambu operator
\begin{equation}
\Psi_{{\bf k}} =
\left(\begin{array}{c}  c_{{\bf k}\uparrow}\\
c_{{\bf k}+{\bf Q}\uparrow} \\
c_{-{\bf k}\downarrow}^{\dag}\\
c_{-{\bf k}-{\bf Q}\downarrow}^{\dag}\\\end{array}\right),
\end{equation}
we can define a $4\times 4$ Hamiltonian matrix $H$ as
\begin{equation}
\Big\langle
\Psi_{{\bf k}}^{\dag}|H|\Psi_{{\bf k}}\Big\rangle =
\begin{centering}\left(\begin{array}{cccc}
(\xi_{{\bf k}}^+ +\xi_{{\bf k}}^-) & -Um & \Delta_{1{\bf k}} &\Delta_{2{\bf k}}\\
-Um & (\xi_{{\bf k}}^+ -\xi_{{\bf k}}^-) & -\Delta_{2{\bf k}} & -\Delta_{1{\bf k}}\\
\Delta^*_{1{\bf k}} & -\Delta^{*}_{2{\bf k}} &-(\xi_{{\bf k}}^+ +\xi_{{\bf k}}^-) &-Um\\
\Delta^{*}_{2{\bf k}} &-\Delta^*_{1{\bf k}}&-Um&-(\xi_{{\bf k}}^+
-\xi_{{\bf k}}^-)
\end{array}\right).
\end{centering}
\end{equation}
We diagonalize the above Hamiltonian in three steps by Bogolyubov transformation method. We  first diagonalize the PDW part, by the following unitary matrix,
\begin{equation}
\left(\begin{array}{c}  c_{{\bf k}\uparrow}\\
c_{{\bf k}+{\bf Q}\uparrow} \\
c_{-{\bf k}\downarrow}^{\dag}\\
c_{-{\bf k}-{\bf Q}\downarrow}^{\dag}\\\end{array}\right)
=\begin{centering}\left(\begin{array}{cccc}
f_{{\bf k}} & 0&0 &-g_{{\bf k}}\\
0&f^*_{{\bf k}}&g^*_{{\bf k}}&0\\
0&-g_{{\bf k}}&f_{{\bf k}}&0\\
g^*_{{\bf k}}&0&0&f^*_{{\bf k}}
\end{array}\right)
\left(\begin{array}{c}  t_{{\bf k}\uparrow}\\
t_{{\bf k}+{\bf Q}\uparrow} \\
t_{-{\bf k}\downarrow}^{\dag}\\
t_{-{\bf k}-{\bf Q}\downarrow}^{\dag}\\\end{array}\right)\end{centering}
\end{equation}
where $f_{\bf k}$ and $g_{\bf k}$ are defined in the main text. 
%
%
SC gap become,
%
%
%
The Hamiltonian then transforms to 
\begin{equation}
H_{\bf k} = 
\begin{centering}\left(\begin{array}{cccc}
( E_{{\bf k}}^{\pi}+\xi_{{\bf k}}^-) & -\Delta_{{\bf k}}^m& \Delta_{{\bf k}}^{\rm SC} &0\\
-\Delta_{{\bf k}}^m& (E_{{\bf k}}^{\pi} -\xi_{{\bf k}}^-) & 0 & -\Delta_{{\bf k}}^{\rm SC}\\
\Delta_{{\bf k}}^{\rm SC} & 0 &-(E_{{\bf k}}^{\pi} +\xi_{{\bf k}}^-) &-\Delta_{{\bf k}}^m\\
0 &-\Delta_{{\bf k}}^{\rm
SC}&-\Delta_{{\bf k}}^m&-(E_{{\bf k}}^{\pi}-\xi_{{\bf k}}^-)
\end{array}\right)\end{centering}
\end{equation}
All quantities here are defined in the main text. The above Hamiltonian looks same as in a AFM-$d$-SC state with
effective AFM- and SC-gaps which is diagonalized by the following unitary matrix\cite{Tanmoytwogap}
\begin{widetext}
\[\left(\begin{array}{c} t_{{\bf k}\uparrow} \\ t_{{\bf k}+{\bf Q}\uparrow}\\
t^{\dag}_{-{\bf k}\downarrow} \\
t^{\dag}_{-{\bf k}-{\bf Q}\downarrow} \\\end{array} \right)=\nonumber\\
\left(\begin{array}{cccc} \alpha_{{\bf k}}u^{+}_{{\bf k}}
&\beta_{{\bf k}}u_{{\bf k}}^{-} & -\alpha_{{\bf k}}
v_{{\bf k}}^{+}
& \beta_{{\bf k}}v^{-}_{{\bf k}}\\
- \beta_{{\bf k}}u^{+}_{{\bf k}} &\alpha_{{\bf k}}u_{{\bf k}}^{-}
& \beta_{{\bf k}}v_{{\bf k}}^{+} &
\alpha_{{\bf k}}v^{-}_{{\bf k}}\\
\alpha_{{\bf k}}v^{+}_{{\bf k}} & \beta_{{\bf k}}v_{{\bf k}}^{-} &
\alpha_{{\bf k}}u_{{\bf k}}^{+} &
-\beta_{{\bf k}} u^{-}_{{\bf k}}\\
\beta_{{\bf k}}v^{+}_{{\bf k}} & -\alpha_{{\bf k}}v_{{\bf k}}^{-}
& \beta_{{\bf k}}u_{{\bf k}}^{+} &
\alpha_{{\bf k}}u^{-}_{{\bf k}}\\\end{array} \right)
\left(\begin{array}{c}B_{{\bf k}\uparrow} \\ B_{{\bf k}+{\bf Q}\uparrow}\\
B^{\dag}_{-{\bf k}\downarrow} \\
B^{\dag}_{-{\bf k}-{\bf Q}\downarrow}
 \\\end{array} \right).\]
\end{widetext}
The AFM cnherences factors $\alpha_{\bf k}$ and $\beta_{\bf k}$ and the SC coherence factors $u_{\bf k}$ and $v_{\bf k}$ are also defined in the main text. Therefore, the total unitary matrix that diagonalizes the full Hamiltonian is
\begin{widetext}
%
\[{\hat U}_{{\bf k}}= \left(\begin{array}{cccc}
f_{{\bf k}}\alpha_{{\bf k}}u_{{\bf k}}^+-g_{{\bf k}}\beta_{{\bf k}}v_{{\bf k}}^+&
f_{{\bf k}}\beta_{{\bf k}}u_{{\bf k}}^-+g_{{\bf k}}\alpha_{{\bf k}}v_{{\bf k}}^-&
-f_{{\bf k}}\alpha_{{\bf k}}v_{{\bf k}}^+-g_{{\bf k}}\beta_{{\bf k}}u_{{\bf k}}^+&
f_{{\bf k}}\beta_{{\bf k}}v_{{\bf k}}^--g_{{\bf k}}\alpha_{{\bf k}}u_{{\bf k}}^-\\
-f^*_{{\bf k}}\beta_{{\bf k}}u_{{\bf k}}^++g^*_{{\bf k}}\alpha_{{\bf k}}v_{{\bf k}}^+&
f^*_{{\bf k}}\alpha_{{\bf k}}u_{{\bf k}}^-+g^*_{{\bf k}}\beta_{{\bf k}}v_{{\bf k}}^-&
f^*_{{\bf k}}\beta_{{\bf k}}v_{{\bf k}}^++g^*_{{\bf k}}\alpha_{{\bf k}}u_{{\bf k}}^+&
f^*_{{\bf k}}\alpha_{{\bf k}}v_{{\bf k}}^--g^*_{{\bf k}}\beta_{{\bf k}}u_{{\bf k}}^-\\
f_{{\bf k}}\alpha_{{\bf k}}v_{{\bf k}}^++g_{{\bf k}}\beta_{{\bf k}}u_{{\bf k}}^+&
f_{{\bf k}}\beta_{{\bf k}}v_{{\bf k}}^--g_{{\bf k}}\alpha_{{\bf k}}u_{{\bf k}}^-&
f_{{\bf k}}\alpha_{{\bf k}}u_{{\bf k}}^+-g_{{\bf k}}\beta_{{\bf k}}v_{{\bf k}}^+&
-f_{{\bf k}}\beta_{{\bf k}}u_{{\bf k}}^--g_{{\bf k}}\alpha_{{\bf k}}v_{{\bf k}}^-\\
f^*_{{\bf k}}\beta_{{\bf k}}v_{{\bf k}}^++g^*_{{\bf k}}\alpha_{{\bf k}}u_{{\bf k}}^+&
-f^*_{{\bf k}}\alpha_{{\bf k}}v_{{\bf k}}^-+g^*_{{\bf k}}\beta_{{\bf k}}u_{{\bf k}}^-&
f^*_{{\bf k}}\beta_{{\bf k}}u_{{\bf k}}^+-g^*_{{\bf k}}\alpha_{{\bf k}}v_{{\bf k}}^+&
f^*_{{\bf k}}\alpha_{{\bf k}}u_{{\bf k}}^-+g^*_{{\bf k}}\beta_{{\bf k}}v_{{\bf k}}^-\\\end{array}\right)\].
\end{widetext}
\section{Self-consistent order parameters}

Now we derive the expression for the self-consistent order parameters in the eigenbasis.

\begin{widetext}
\begin{eqnarray}\label{td}
\Delta_{10}&=&V_{10}\sum_{{\bf k}}d_{{\bf k}}\Big\langle c^{\dag}_{{\bf k}\uparrow}c^{\dag}_{-{\bf k}\downarrow}\Big\rangle = V_{10}\sum^{\prime}_{{\bf k}}d_{{\bf k}}\Big[\Big\langle
c^{\dag}_{{\bf k}\uparrow}c^{\dag}_{-{\bf k}\downarrow}\Big\rangle
-\Big\langle c^{\dag}_{{\bf k}+{\bf Q}\uparrow}c^{\dag}_{-{\bf k}-{\bf Q}\downarrow}\Big\rangle \Big]\nonumber\\
&=&V_{10}\sum^{\prime}_{{\bf k}}d_{{\bf k}}
\Big[\Big((f_{{\bf k}}\alpha_{{\bf k}}u_{{\bf k}}^+-g_{{\bf k}}\beta_{{\bf k}}v_{{\bf k}}^+)(f_{{\bf k}}\alpha_{{\bf k}}v_{{\bf k}}^++g_{{\bf k}}\beta_{{\bf k}}u_{{\bf k}}^+)
\big\langle B^{\dag}_{{\bf k},\uparrow}B_{{\bf k},\uparrow}\big\rangle\nonumber\\
&&~~+(f_{{\bf k}}\beta_{{\bf k}}u_{{\bf k}}^-+g_{{\bf k}}\alpha_{{\bf k}}v_{{\bf k}}^-)(f_{{\bf k}}\beta_{{\bf k}}v_{{\bf k}}^--g_{{\bf k}}\alpha_{{\bf k}}u_{{\bf k}}^-)
\big\langle B^{\dag}_{{\bf k}+{{\bf Q}},\uparrow}B_{{\bf k}+{{\bf Q}},\uparrow}\big\rangle\nonumber\\
&&~~+(-f_{{\bf k}}\alpha_{{\bf k}}v_{{\bf k}}^+-g_{{\bf k}}\beta_{{\bf k}}u_{{\bf k}}^+)(f_{{\bf k}}\alpha_{{\bf k}}u_{{\bf k}}^+-g_{{\bf k}}\beta_{{\bf k}}v_{{\bf k}}^+)
\big\langle B^{\dag}_{-{\bf k},\uparrow}B_{-{\bf k},\uparrow}\big\rangle\nonumber\\
&&~~+(f_{{\bf k}}\beta_{{\bf k}}v_{{\bf k}}^--g_{{\bf k}}\alpha_{{\bf k}}u_{{\bf k}}^-)(-f_{{\bf k}}\beta_{{\bf k}}u_{{\bf k}}^--g_{{\bf k}}\alpha_{{\bf k}}v_{{\bf k}}^-)
\big\langle B^{\dag}_{-{\bf k}+{{\bf Q}},\uparrow}B_{-{\bf k}+{{\bf Q}},\uparrow}\Big\rangle\Big)\nonumber\\
&&-(-f^*_{{\bf k}}\beta_{{\bf k}}u_{{\bf k}}^++g^*_{{\bf k}}\alpha_{{\bf k}}v_{{\bf k}}^+)(f^*_{{\bf k}}\beta_{{\bf k}}v_{{\bf k}}^++g^*_{{\bf k}}\alpha_{{\bf k}}u_{{\bf k}}^+)
\big\langle B^{\dag}_{{\bf k},\uparrow}B_{{\bf k},\uparrow}\big\rangle\nonumber\\
&&~~-(f^*_{{\bf k}}\alpha_{{\bf k}}u_{{\bf k}}^-+g^*_{{\bf k}}\beta_{{\bf k}}v_{{\bf k}}^-)(-f^*_{{\bf k}}\alpha_{{\bf k}}v_{{\bf k}}^-+g^*_{{\bf k}}\beta_{{\bf k}}u_{{\bf k}}^-)
\big\langle B^{\dag}_{{\bf k}+{{\bf Q}},\uparrow}B_{{\bf k}+{{\bf Q}},\uparrow}\big\rangle\nonumber\\
&&~~-(f^*_{{\bf k}}\beta_{{\bf k}}v_{{\bf k}}^++g^*_{{\bf k}}\alpha_{{\bf k}}u_{{\bf k}}^+)(f^*_{{\bf k}}\beta_{{\bf k}}u_{{\bf k}}^+-g^*_{{\bf k}}\alpha_{{\bf k}}v_{{\bf k}}^+)
\big\langle B^{\dag}_{-{\bf k},\uparrow}B_{-{\bf k},\uparrow}\big\rangle\nonumber\\
&&~~-(f^*_{{\bf k}}\alpha_{{\bf k}}v_{{\bf k}}^--g^*_{{\bf k}}\beta_{{\bf k}}u_{{\bf k}}^-)(f^*_{{\bf k}}\alpha_{{\bf k}}u_{{\bf k}}^-+g^*_{{\bf k}}\beta_{{\bf k}}v_{{\bf k}}^-)\big\langle B^{\dag}_{-{\bf k}+{{\bf  Q}},\uparrow}B_{-{\bf k}+{{\bf Q}},\uparrow}\Big\rangle\Big]\nonumber\\
&=&V_{10}\sum^{\prime}_{{\bf k}}\sum_{\nu=\pm}d_{{\bf k}}\Big[{\rm Re}[f^2_{{\bf k}}-g^2_{{\bf k}}]u_{{\bf k}}^{\nu}v_{{\bf k}}^{\nu}+2{\rm Re}[f_{{\bf k}}g_{{\bf k}}]\alpha_{{\bf k}}\beta_{{\bf k}}((u^{\nu}_{{\bf k}})^2-(v^{\nu}_{{\bf k}})^2)\Big]\tanh{(\beta
E_{{\bf k}}^{{\nu}}/2)}
\end{eqnarray}
\end{widetext}
In the first line and right hand side of the above equation we have taken into account of the fact that the $d$-wave structure factor $s_{{{\bf k}}+{{\bf Q}}}=-s_{{\bf k}}$ for this particular commensurate wavevector. Since we have identified $E_{{\bf k}}^{\pm}$ as the excitation energies of the fermion quasi-particles, the probability of its excitation in thermal equilibrium is the usual Fermi function,
\begin{equation}\label{fermifunction}
\langle
B_{{\bf k}\sigma}^{\dag}B_{{\bf k}\sigma}\rangle=n(E_{{\bf k}}^{+})
= (\exp{(\beta E_{{\bf k}}^{+})}+1)^{-1},
\end{equation}
with $\beta=1/k_BT$ and so on. Therefore
\begin{eqnarray}\label{expectation}
\langle 1- B_{{\bf k}\sigma}^{\dag}B_{{\bf k}\sigma} -
B_{-{\bf k}\bar{\sigma}}^{\dag}B_{-{\bf k}\bar{\sigma}}\rangle
&=& 1-2n(E_{{\bf k}}^{+})\nonumber\\
&=&\tanh{(\beta E_{{\bf k}}^{+}/2)}.
\end{eqnarray}
The PDW SC order is
\begin{eqnarray}\label{tt}
\Delta_{20}&=&V_{20}\sum_{{\bf k}}p_{{\bf k}}
\Big\langle c^{\dag}_{{\bf k}\uparrow}c^{\dag}_{-{\bf k}-{\bf Q}\downarrow}\Big\rangle.
\end{eqnarray}
As mentioned before, the difference between the PDW and singlet pairings in the magnetic Brillouin zone is that in the former case, the pairing symmetry picks up the complex conjugate as we go from the main band to the AFM subbands shifted by ${{\bf Q}}$ vector, and also changes sign. Writing in Bogolyubov quasiparticle form as before, we get
\begin{widetext}
\begin{eqnarray}\label{tt}
\Delta_{20}&=&V_{20}\sum^{\prime}_{{\bf k}}\Big[p_{{\bf k}}\big((f_{{\bf k}}\alpha_{{\bf k}}u_{{\bf k}}^+-g_{{\bf k}}\beta_{{\bf k}}v_{{\bf k}}^+)(f^*_{{\bf k}}\beta_{{\bf k}}v_{{\bf k}}^++g^*_{{\bf k}}\alpha_{{\bf k}}u_{{\bf k}}^+)
\big\langle B^{\dag}_{{\bf k},\uparrow}B_{{\bf k},\uparrow}\big\rangle\nonumber\\
&&~~+(f_{{\bf k}}\beta_{{\bf k}}u_{{\bf k}}^-+g_{{\bf k}}\alpha_{{\bf k}}v_{{\bf k}}^-)(-f^*_{{\bf k}}\alpha_{{\bf k}}v_{{\bf k}}^-+g^*_{{\bf k}}\beta_{{\bf k}}u_{{\bf k}}^-)\big\langle B^{\dag}_{{\bf k}+{{\bf  Q}},\uparrow}B_{{\bf k}+{{\bf Q}},\uparrow}\big\rangle\nonumber\\
&&~~+(-f_{{\bf k}}\alpha_{{\bf k}}v_{{\bf k}}^+-g_{{\bf k}}\beta_{{\bf k}}u_{{\bf k}}^+)(f^*_{{\bf k}}\beta_{{\bf k}}u_{{\bf k}}^+-g^*_{{\bf k}}\alpha_{{\bf k}}v_{{\bf k}}^+)\big\langle B^{\dag}_{-{\bf k},\uparrow}B_{-{\bf k},\uparrow}\big\rangle\nonumber\\
&&~~+(f_{{\bf k}}\beta_{{\bf k}}v_{{\bf k}}^--g_{{\bf k}}\alpha_{{\bf k}}u_{{\bf k}}^-)(f^*_{{\bf k}}\alpha_{{\bf k}}u_{{\bf k}}^-+g^*_{{\bf k}}\beta_{{\bf k}}v_{{\bf k}}^-)\big\langle B^{\dag}_{-{\bf k}+{{\bf  Q}},\uparrow}B_{-{\bf k}+{{\bf Q}},\uparrow}\Big\rangle\Big)\nonumber\\
&&- p^*_{{\bf k}}\Big((-f^*_{{\bf k}}\beta_{{\bf k}}u_{{\bf k}}^++g^*_{{\bf k}}\alpha_{{\bf k}}v_{{\bf k}}^+)(f_{{\bf k}}\alpha_{{\bf k}}v_{{\bf k}}^++g_{{\bf k}}\beta_{{\bf k}}u_{{\bf k}}^+)\big\langle B^{\dag}_{{\bf k},\uparrow}B_{{\bf k},\uparrow}\big\rangle\nonumber\\
&&~~+(f^*_{{\bf k}}\alpha_{{\bf k}}u_{{\bf k}}^-+g^*_{{\bf k}}\beta_{{\bf k}}v_{{\bf k}}^-)(f_{{\bf k}}\beta_{{\bf k}}v_{{\bf k}}^--g_{{\bf k}}\alpha_{{\bf k}}u_{{\bf k}}^-)\big\langle B^{\dag}_{{\bf k}+{{\bf  Q}},\uparrow}B_{{\bf k}+{{\bf Q}},\uparrow}\big\rangle\nonumber\\
&&~~+(f^*_{{\bf k}}\beta_{{\bf k}}v_{{\bf k}}^++g^*_{{\bf k}}\alpha_{{\bf k}}u_{{\bf k}}^+)(f_{{\bf k}}\alpha_{{\bf k}}u_{{\bf k}}^+-g_{{\bf k}}\beta_{{\bf k}}v_{{\bf k}}^+)\big\langle B^{\dag}_{-{\bf k},\uparrow}B_{-{\bf k},\uparrow}\big\rangle\nonumber\\
&&~~+(f^*_{{\bf k}}\alpha_{{\bf k}}v_{{\bf k}}^--g^*_{{\bf k}}\beta_{{\bf k}}u_{{\bf k}}^-)(-f_{{\bf k}}\beta_{{\bf k}}u_{{\bf k}}^--g_{{\bf k}}\alpha_{{\bf k}}v_{{\bf k}}^-)
\big\langle B^{\dag}_{-{\bf k}+{{\bf Q}},\uparrow}B_{-{\bf k}+{{\bf Q}},\uparrow}\Big\rangle
\Big)\Big]\nonumber\\
&=&V_{20}\sum^{\prime}_{{\bf k}}p_{{\bf k}}\sum_{\nu}\Big[\big[{\rm Re}[f_{{\bf k}}g_{{\bf k}}]((u^{\nu}_{{\bf k}})^2-(v^{\nu}_{{\bf k}})^2)+2\alpha_{{\bf k}}\beta_{{\bf k}}u_{{\bf k}}^{\nu}v_{{\bf k}}^{\nu}(|f_{{\bf k}}|^2-|g_{{\bf k}}|^2)\big]\tanh{(\beta
E_{{\bf k}}^{\nu}/2)})\Big].
\end{eqnarray}
Finally, the staggered magnetic is
\begin{eqnarray}\label{ts}
m&=&\sum_{{\bf k},\sigma}\sigma\Big\langle
c^{\dag}_{{\bf k}+{\bf Q}\sigma}c_{{\bf k}\sigma}\Big\rangle
=\sum_{{\bf k}}\Big[\Big\langle
c^{\dag}_{{\bf k}+{\bf Q}\uparrow}c_{{\bf k}\uparrow}\Big\rangle-\Big\langle
c^{\dag}_{-{\bf k}-{\bf Q}\downarrow}c_{-{\bf k}\downarrow}\Big\rangle\Big]\nonumber\\
&=&\sum^{\prime}_{{\bf k}}\Big[\Big\langle
c^{\dag}_{{\bf k}+{\bf Q}\uparrow}c_{{\bf k}\uparrow}\Big\rangle
+\Big\langle
c^{\dag}_{{\bf k}\uparrow}c_{{\bf k}+{\bf Q}\uparrow}\Big\rangle
-\Big\langle
c^{\dag}_{-{\bf k}-{\bf Q}\downarrow}c_{-{\bf k}\downarrow}\Big\rangle
-\Big\langle
c^{\dag}_{-{\bf k}\downarrow}c_{-{\bf k}-{\bf Q}\downarrow}\Big\rangle\Big]\nonumber\\
&=&\sum^{\prime}_{{\bf k}}\Big[(f_{{\bf k}}\alpha_{{\bf k}}u_{{\bf k}}^+-g_{{\bf k}}\beta_{{\bf k}}v_{{\bf k}}^+)(-f^*_{{\bf k}}\beta_{{\bf k}}u_{{\bf k}}^++g^*_{{\bf k}}\alpha_{{\bf k}}v_{{\bf k}}^+)
\big\langle B^{\dag}_{{\bf k},\uparrow}B_{{\bf k},\uparrow}\big\rangle\nonumber\\
&&~~~+(f_{{\bf k}}\beta_{{\bf k}}u_{{\bf k}}^-+g_{{\bf k}}\alpha_{{\bf k}}v_{{\bf k}}^-)(f^*_{{\bf k}}\alpha_{{\bf k}}u_{{\bf k}}^-+g^*_{{\bf k}}\beta_{{\bf k}}v_{{\bf k}}^-)
\big\langle B^{\dag}_{{\bf k}+{{\bf Q}},\uparrow}B_{{\bf k}+{{\bf Q}},\uparrow}\big\rangle\nonumber\\
&&~~~+(-f_{{\bf k}}\alpha_{{\bf k}}v_{{\bf k}}^+-g_{{\bf k}}\beta_{{\bf k}}u_{{\bf k}}^+)(f^*_{{\bf k}}\beta_{{\bf k}}v_{{\bf k}}^++g^*_{{\bf k}}\alpha_{{\bf k}}u_{{\bf k}}^+)
\big\langle B^{\dag}_{-{\bf k},\uparrow}B_{-{\bf k},\uparrow}\big\rangle\nonumber\\
&&~~~+(f_{{\bf k}}\beta_{{\bf k}}v_{{\bf k}}^--g_{{\bf k}}\alpha_{{\bf k}}u_{{\bf k}}^-)(f^*_{{\bf k}}\alpha_{{\bf k}}v_{{\bf k}}^--g^*_{{\bf k}}\beta_{{\bf k}}u_{{\bf k}}^-)
\big\langle B^{\dag}_{-{\bf k}+{{\bf Q}},\uparrow}B_{-{\bf k}+{{\bf Q}},\uparrow}\Big\rangle\nonumber\\
&&~~~-(f_{{\bf k}}\alpha_{{\bf k}}v_{{\bf k}}^++g_{{\bf k}}\beta_{{\bf k}}u_{{\bf k}}^+)(f^*_{{\bf k}}\beta_{{\bf k}}v_{{\bf k}}^++g^*_{{\bf k}}\alpha_{{\bf k}}u_{{\bf k}}^+)
\big\langle B^{\dag}_{{\bf k},\uparrow}B_{{\bf k},\uparrow}\big\rangle
\nonumber\\
&&~~~-(f_{{\bf k}}\beta_{{\bf k}}v_{{\bf k}}^--g_{{\bf k}}\alpha_{{\bf k}}u_{{\bf k}}^-)(-f^*_{{\bf k}}\alpha_{{\bf k}}v_{{\bf k}}^-+g^*_{{\bf k}}\beta_{{\bf k}}u_{{\bf k}}^-)
\big\langle B^{\dag}_{{\bf k}+{{\bf Q}},\uparrow}B_{{\bf k}+{{\bf Q}},\uparrow}\big\rangle\nonumber\\
&&~~~-(f_{{\bf k}}\alpha_{{\bf k}}u_{{\bf k}}^+-g_{{\bf k}}\beta_{{\bf k}}v_{{\bf k}}^+)(f^*_{{\bf k}}\beta_{{\bf k}}u_{{\bf k}}^+-g^*_{{\bf k}}\alpha_{{\bf k}}v_{{\bf k}}^+)
\big\langle B^{\dag}_{-{\bf k},\uparrow}B_{-{\bf k},\uparrow}\big\rangle\nonumber\\
&&~~~-(-f_{{\bf k}}\beta_{{\bf k}}u_{{\bf k}}^--g_{{\bf k}}\alpha_{{\bf k}}v_{{\bf k}}^-)(f^*_{{\bf k}}\alpha_{{\bf k}}u_{{\bf k}}^-+g^*_{{\bf k}}\beta_{{\bf k}}v_{{\bf k}}^-)
\big\langle B^{\dag}_{-{\bf k}+{{\bf Q}},\uparrow}B_{-{\bf k}+{{\bf Q}},\uparrow}\Big\rangle\Big],\nonumber\\
&=&\sum^{\prime}_{{\bf k}}\alpha_{{\bf k}}\beta_{{\bf k}}(|f_{{\bf k}}|^2-|g_{{\bf k}}|^2)
\Big[((v_{{\bf k}}^{-})^2-(v_{{\bf k}}^{-})^2)
+((v_{{\bf k}}^{+})^2-(u_{{\bf k}}^{+})^2)n(E_{{\bf k}}^{+})-((v_{{\bf k}}^{-})^2-(u_{{\bf k}}^{-})^2)n(E_{{\bf k}}^{-})\Big].
\end{eqnarray}
\end{widetext}

\section{Penetration Depth}

The quantum mechanical electric current can be written as
\begin{equation}\label{jnoA}
{\bf J}=-\frac{e}{2m}(\psi^{\dag}{\bf p}\psi-({\bf p}\psi)^{\dag}\psi)
\end{equation}
Now in a magnetic field the momentum operator becomes
${\bf p}+e/c{\bf A}$, so that the current is
\begin{eqnarray}\label{jA1}
{\bf J}&=&-\frac{e}{2m}(\psi^{\dag}({\bf p}+e/c{\bf A})\psi-[({\bf p}+e/c\bf(A))\psi]^{\dag}\psi)\nonumber\\
&=&-\frac{e}{2m}(\psi^{\dag}{\bf p}\psi-({\bf p}\psi)^{\dag}\psi)-\frac{e^2{\bf A}}{mc}\psi^{\dag}\psi
\end{eqnarray}
The first term represents the current due to the normal electrons
(paramagnetic current) and the second term is identified as the
diamagnetic current which is contributed by the superconducting
electrons $n_s$. The paramagnetic current (${\bf J}_n$) has the
tendency to cancel the diamagnetic current (${\bf J}_s$).
Therefore the velocity of the superconducting electrons are
identified as ${\bf v}_s^{*}=\frac{e{\bf A}}{m^{*}c}$ where $m^{*}$, the band mass, is a $2\times2$ tensor in two
dimensional space defined below.

The Fourier transformation of these quantities gives

\begin{eqnarray}\label{fouriertrans}
{\bf J}({\bf q})&=&\int
{\bf J}({{\bf r}})e^{-i{\bf q}.{\bf r}}d^3{\bf r},\nonumber\\
{\bf a}({\bf q})&=&\int
{\bf A}({{\bf r}})e^{-i{\bf q}.{\bf r}}d^3{\bf r},\nonumber\\
{\bf K}({\bf q})&=&\int
{\bf K}({{\bf r}})e^{-i{\bf q}.{\bf r}}d^3{\bf r},
\end{eqnarray}
which gives
\begin{eqnarray}\label{JAK}
{\bf J}({\bf q})&=&{\bf J_n}({\bf q})+{\bf J_s}({\bf q})
={\bf A}({\bf q}).\overrightarrow{\overleftarrow{K}}({\bf q}),
\end{eqnarray}
where ${\bf K}({\bf q})$ is called the response function. Now in
an anisotropic system the response function becomes a $2\times2$
tensor as defined below
\begin{eqnarray}
\left(\begin{array}{c} J_x({\bf q})\\J_y({\bf q}) \\\end{array} \right)&=&\left(\begin{array}{c} J_{nx}({\bf q})+J_{sx}({\bf q})\\J_{ny}({\bf q})+J_{sy}({\bf q}) \\\end{array} \right)\nonumber\\
&=& -\frac{c}{4\pi}\left(\begin{array}{cc} K_{xx}({\bf q}) & K_{xy}({\bf q}) \\
K_{yx}({\bf q})& K_{yy}({\bf q})\\\end{array}\right)
 \left(\begin{array}{c} a_x({\bf q})\\a_y({\bf q}) \\\end{array} \right)\nonumber\\
\end{eqnarray}
where at ${\bf q}=0$, the penetration depth is related to $K$s as
\begin{equation}\label{Klambda}
 \lambda_{ij}^{-2}(T)=K_{ij}(0)
 \end{equation}
\subsection{Paramagnetic Current, ${\bf J}_p({\bf q})$}
The first term ${\bf J_p}$ arises due to normal electrons, which is often called the ``paramagnetic current" term because it tends to cancel the diamagnetic current ${\bf J_d}$, which aries because of
superconducting electrons. In the presence of an electromagnetic field (the field includes the effects of screening supercurrents), the canonical momentum is modified and the kinetic energy is $({\bf p^{*}}-e{\bf A}/c)^2/2m^*$, where ${\bf p}^{*}$ is the crystal momentum or the band momentum. Thus the resulting
perturbation Hamiltonian term is
\begin{eqnarray}\label{intH3}
H_I& =&-\frac{e}{2c}\sum_i\frac{1}{m^*}({\bf p}^{*}.{\bf A} +
{\bf A}.{\bf p}^{*})\nonumber\\
&=&-\frac{e}{c}\sum_i{\bf v}^{*}.{\bf A}
\end{eqnarray}
The band momentum is defined as ${\bf p}^{*} = m^{*}{\bf v}^{*}$,
where ${\bf v}^{*}$ is the band velocity for the band $\xi_{{\bf k}}^{+}$. Note that the symbol $v_{{\bf k}}$ without a superscript of $\pm$ is the band velocity, while with the superscript it gives the AFM conherence factors.

\begin{widetext}
\begin{eqnarray}\label{tPenIntHam}
{\bf J}_p(0)&=&-e{\bf v}^*n\nonumber\\
&=&-\frac{e}{\Omega}\sum_{{\bf k}}^{\prime}\Big[{\bf v}_{{\bf k}}\big\langle c_{{\bf k},\uparrow}^{\dag}c_{{\bf k},\uparrow}\big\rangle
+{\bf v}_{{\bf k}+{\bf Q}}\big\langle c_{{\bf k}+{\bf Q},\uparrow}^{\dag}c_{{\bf k}+{\bf Q},\uparrow}\big\rangle
-{\bf v}_{{\bf k}}\big\langle c_{-{\bf k},\downarrow}^{\dag}c_{-{\bf k},\downarrow}\big\rangle
-{\bf v}_{{\bf k}+{\bf Q}}\big\langle c_{-{\bf k}-{\bf Q},\downarrow}^{\dag}c_{-{\bf k}-{\bf Q},\downarrow}\big\rangle\Big]\nonumber\\
&=&-\frac{e}{c\Omega}{\bf a}(0)\sum_{{\bf k}}^{\prime}\Big[
f_{{\bf k}}^2\Big({\bf v}_{{\bf k}}\alpha_{{\bf k}}^2+{\bf v}_{{\bf k}+{\bf Q}}\beta_{{\bf k}}^2\Big)
+g_{{\bf k}}^2\Big({\bf v}_{{\bf k}}\beta_{{\bf k}}^2+{\bf v}_{{\bf k}+{\bf Q}}\alpha_{{\bf k}}^2\Big)\Big]
\Big(\Big\langle B_{{\bf k},\uparrow}^{\dag}B_{{\bf k},\uparrow}\Big\rangle-\Big\langle B_{-{\bf k},\downarrow}^{\dag}B_{-{\bf k},\downarrow}\Big\rangle\Big)\nonumber\\
&&~~~+\Big[
f_{{\bf k}}^2\Big({\bf v}_{{\bf k}}\beta_{{\bf k}}^2+{\bf v}_{{\bf k}+{\bf Q}}\alpha_{{\bf k}}^2\Big)
+g_{{\bf k}}^2\Big({\bf v}_{{\bf k}}\alpha_{{\bf k}}^2+{\bf v}_{{\bf k}+{\bf Q}}\beta_{{\bf k}}^2\Big)\Big]
\Big(\Big\langle B_{{\bf k}+{\bf Q},\uparrow}^{\dag}B_{{\bf k}+{\bf Q},\uparrow}\Big\rangle-\Big\langle B_{-{\bf k}-{\bf Q},\downarrow}^{\dag}B_{-{\bf k}-{\bf Q},\downarrow}\Big\rangle\Big)\nonumber\\
&=&-\frac{e}{c\Omega}{\bf a}(0)\sum_{{\bf k}}^{\prime}{\bf V}_{{\bf k}}^+
\Big(\big\langle B_{{\bf k},\uparrow}^{\dag}B_{{\bf k},\uparrow}\big\rangle-\big\langle B_{-{\bf k},\downarrow}^{\dag}B_{-{\bf k},\downarrow}\big\rangle\Big)
+{\bf V}_{{\bf k}}^-
\Big(\big\langle B_{{\bf k}+{\bf Q},\uparrow}^{\dag}B_{{\bf k}+{\bf Q},\uparrow}\big\rangle-\big\langle B_{-{\bf k}-{\bf Q},\downarrow}^{\dag}B_{-{\bf k}-{\bf Q},\downarrow}\big\rangle\Big).
\end{eqnarray}
\end{widetext}
Here $\Omega$ is the unit cell volume. ${\bf V}_{{\bf k}}^+=C_{{\bf k}}{\bf v}_{{\bf k}} + C_{{\bf k}+{\bf Q}}{\bf v}_{{\bf k}+{\bf Q}} $ and ${\bf V}_{{\bf k}}^-={\bf V}_{{\bf k}+{\bf Q}}^+$, and $C_{{\bf k}}=|f_{{\bf k}}|^2\alpha^2_{{\bf k}}+|g_{{\bf k}}^2\beta^2_{{\bf k}}$. Now the probability of excitation of the fermion quasiparticles in an electromagnetic field in the thermal equilibrium is the
usual Fermi function in the corresponding energy level sifted by the electromagnetic field. In the limit of small ${\bf a}(0)$, we can expand the Fermi function in
Taylor's series and keeping only the first term we get,
\begin{eqnarray}\label{taylor}
n(E_{{\bf k}\pm}^{\nu}) &=& n(E_{{\bf k}}^{\nu} \mp \frac{e}{c}{\bf v}_{{\bf k}}^{\nu}.{\bf a}(0))\nonumber\\
&\approx& n( E_{{\bf k}}^{\nu}) \mp \frac{e}{c}\left(\frac{\partial n(E_{{\bf k}}^{\nu})}{\partial E_{{\bf k}}^{\nu}}\right){\bf V}_{{\bf k}}^{\nu}.{\bf a}(0).
\end{eqnarray}

Here $\nu=\pm$ are two quasiparticle bands. Then adding and subtracting the fermi functions we have
\begin{eqnarray}
n(E_{{\bf k}+}^{\nu})+n(E_{{\bf k}-}^{\nu}) &\approx&
2n(E_{{\bf k}}^{\nu})\label{diffn}\\
n(E_{{\bf k}+}^{\nu})-n(E_{{\bf k}-}^{\nu}) &\approx&
-\frac{2e}{c}\left(\frac{\partial n(E_{{\bf k}}^{\nu})}{\partial
E_{{\bf k}}^{\nu}}\right){\bf V}_{{\bf k}}^{\nu}.{\bf a}(0)\label{diffp}
\end{eqnarray}

Substituting Eq.~\ref{diffp} in Eq.~\ref{tPenIntHam}, we get the paramagnetic current as
\begin{eqnarray}\label{Jp}
{\bf J}_p(0)&=&-\frac{2e^2}{c\Omega}\sum_{{\bf k},\nu}^{\prime}
{\bf V}^{\nu}\Big({\bf V}^{\nu}.{\bf a}(0)\Big)\left(-\frac{\partial
f(E^{\nu}_{{\bf k}})}{\partial E^{\nu}_{{\bf k}}}\right)\nonumber\\
&=&-\frac{e^2}{c\Omega}\frac{\beta}{2}\sum_{{\bf k},\nu}^{\prime}
{\bf V}^{\nu}\Big({\bf V}^{\nu}.{\bf a}(0)\Big){\rm sech}^2(\beta
E^{\nu}_{{\bf k}}/2)
\end{eqnarray}
\subsection{Diamagnetic Current}
Let us define
\begin{eqnarray}\label{tbandmass}
\frac{1}{M^+_{{\bf k}i,j}}&=&C_{{\bf k}}\left(\frac{1}{m_{{\bf k}i,j}}\right) + C_{{\bf k}+{\bf Q}}\left(\frac{1}{m_{{\bf k}+{\bf Q}i,j}}\right),\nonumber\\
\frac{1}{m^{\nu}_{{\bf k}i,j}}&=&4f_{{\bf k}}g_{{\bf k}}\alpha_{{\bf k}}\beta_{{\bf k}}u_{{\bf k}}^{\nu}v_{{\bf k}}^{\nu}\left(\frac{1}{m^*_{{\bf k}i,j}}+\frac{1}{m^*_{{\bf k}+{\bf Q}i,j}}\right),\nonumber\\
\end{eqnarray}
and $\frac{1}{M^-_{{\bf k}i,j}}=\frac{1}{M^+_{{\bf k}+{\bf Q}i,j}}$. Then the $i^{th}$ component of the diamagnetic current is
\begin{widetext}
\begin{eqnarray}\label{Jd}
J_{di}(0)&=&-\frac{e^2}{c}\sum_{j=1}^2\left(\frac{1}{m^*_{i,j}}\right)a_j(0)n\nonumber\\
&=&-\frac{e^2}{c\Omega}\sum_{j=1}^2a_j(0)\sum_{{\bf k}}^{\prime}\Big[\left(\frac{1}{m^*_{{\bf k}i,j}}\right)\big\langle c_{{\bf k},\uparrow}^{\dag}c_{{\bf k},\uparrow}\big\rangle
+\left(\frac{1}{m^*_{{\bf k}+{\bf Q}i,j}}\right)\big\langle c_{{\bf k}+{\bf Q},\uparrow}^{\dag}c_{{\bf k}+{\bf Q},\uparrow}\big\rangle\nonumber\\
&&~~~~~~~~~~~~~~~~~~~+\left(\frac{1}{m^*_{{\bf k}i,j}}\right)\big\langle c_{-{\bf k},\downarrow}^{\dag}c_{-{\bf k},\downarrow}\big\rangle
+\left(\frac{1}{m^*_{{\bf k}+{\bf Q}i,j}}\right)\big\langle c_{-{\bf k}-{\bf Q},\downarrow}^{\dag}c_{-{\bf k}-{\bf Q},\downarrow}\big\rangle\Big]\nonumber\\
&=&-\frac{e^2}{c\Omega}\sum_{j=1}^2a_j(0)\sum_{{\bf k}}^{\prime}\left[\left(\frac{1}{m^*_{i,j}}\right)\Big\{
(f_{{\bf k}}\alpha_{{\bf k}}u_{{\bf k}}^+-g_{{\bf k}}\beta_{{\bf k}}v_{{\bf k}}^+)^2
-(f_{{\bf k}}\alpha_{{\bf k}}v_{{\bf k}}^++g_{{\bf k}}\beta_{{\bf k}}u_{{\bf k}}^+)^2\Big\}\right.\nonumber\\
&&~~~~~~~~~~~~+\left.\left(\frac{1}{m^*_{{\bf k}+{\bf Q}i,j}}\right)\Big\{
(f_{{\bf k}}\beta_{{\bf k}}u_{{\bf k}}^+-g_{{\bf k}}\alpha_{{\bf k}}v_{{\bf k}}^+)^2
-(f_{{\bf k}}\beta_{{\bf k}}v_{{\bf k}}^++g_{{\bf k}}\alpha_{{\bf k}}u_{{\bf k}}^+)^2\Big\}\right]
\nonumber\\
&&\qquad\qquad\times\left(\Big\langle B_{{\bf k},\uparrow}^{\dag}B_{{\bf k},\uparrow}\Big\rangle
+\Big\langle B_{-{\bf k},\downarrow}^{\dag}B_{-{\bf k},\downarrow}\Big\rangle\right) \nonumber\\
&&~~~~~~~~~+2\left(\frac{1}{m^*_{i,j}}\right)(f_{{\bf k}}\alpha_{{\bf k}}v_{{\bf k}}^++g_{{\bf k}}\beta_{{\bf k}}u_{{\bf k}}^+)^2
+2\left(\frac{1}{m^*_{{\bf k}+{\bf Q}i,j}}\right){\bf v}_{{\bf k}+{\bf Q}}(f_{{\bf k}}\beta_{{\bf k}}v_{{\bf k}}^++g_{{\bf k}}\alpha_{{\bf k}}u_{{\bf k}}^+)^2\nonumber\\
&&~~~~~~~~~~~~+\left[\left(\frac{1}{m^*_{i,j}}\right)\Big\{
(f_{{\bf k}}\beta_{{\bf k}}u_{{\bf k}}^-+g_{{\bf k}}\alpha_{{\bf k}}v_{{\bf k}}^-)^2
-\left(\frac{1}{m^*_{i,j}}\right)(f_{{\bf k}}\beta_{{\bf k}}v_{{\bf k}}^--g_{{\bf k}}\alpha_{{\bf k}}u_{{\bf k}}^-)^2\Big\}\right.\nonumber\\
&&~~~~~~~~~~~~\left.+\left(\frac{1}{m^*_{{\bf k}+{\bf Q}i,j}}\right)\Big\{
(f_{{\bf k}}\alpha_{{\bf k}}u_{{\bf k}}^-+g_{{\bf k}}\beta_{{\bf k}}v_{{\bf k}}^-)^2
-(f_{{\bf k}}\alpha_{{\bf k}}v_{{\bf k}}^--g_{{\bf k}}\beta_{{\bf k}}u_{{\bf k}}^-)^2\Big\}\right]\nonumber\\
&&\qquad\qquad\left(\Big\langle
B_{{\bf k}+{\bf Q},\uparrow}^{\dag}B_{{\bf k},\uparrow}\Big\rangle
+\Big\langle B_{-{\bf k}-{\bf Q},\downarrow}^{\dag}B_{-{\bf k},\downarrow}\Big\rangle\right)\nonumber\\
&&~~~~~~~~~+2\left(\frac{1}{m^*_{i,j}}\right)(f_{{\bf k}}\beta_{{\bf k}}v_{{\bf k}}^--g_{{\bf k}}\alpha_{{\bf k}}u_{{\bf k}}^-)^2
+2\left(\frac{1}{m^*_{{\bf k}+{\bf Q}i,j}}\right)(f_{{\bf k}}\alpha_{{\bf k}}v_{{\bf k}}^--g_{{\bf k}}\beta_{{\bf k}}u_{{\bf k}}^-)^2,\nonumber\\
\end{eqnarray}
\begin{eqnarray}\label{Jd}
J_{di}(0)&=&-\frac{e^2}{c\Omega}\sum_{j=1}^2a_j(0)\sum_{{\bf k}}^{\prime} f_{{\bf k}}^2\left(\left(\frac{1}{m^*_{i,j}}\right)\alpha_{{\bf k}}^2
+\left(\frac{1}{m^*_{{\bf k}+{\bf Q}i,j}}\right)\beta_{{\bf k}}^2\right)\nonumber\\
&&\qquad\times \left[2(v_{{\bf k}}^+)^2+\Big((u_{{\bf k}}^+)^2-(v_{{\bf k}}^+)^2\Big)\Big(1-\Big\langle
B_{{\bf k},\uparrow}^{\dag}B_{{\bf k},\uparrow}\Big\rangle
-\Big\langle
B_{-{\bf k},\downarrow}^{\dag}B_{-{\bf k},\downarrow}\Big\rangle\Big)\right]\nonumber\\
&&+g_{{\bf k}}^2\left(\left(\frac{1}{m^*_{i,j}}\right)\beta_{{\bf k}}^2
+\left(\frac{1}{m^*_{{\bf k}+{\bf Q}i,j}}\right)\alpha_{{\bf k}}^2\right)\nonumber\\
&&\qquad\times\left[2(u_{{\bf k}}^+)^2-\Big((u_{{\bf k}}^+)^2-(v_{{\bf k}}^+)^2\Big)
\Big(1-\Big\langle
B_{{\bf k},\uparrow}^{\dag}B_{{\bf k},\uparrow}\Big\rangle
-\Big\langle
B_{-{\bf k},\downarrow}^{\dag}B_{-{\bf k},\downarrow}\Big\rangle\Big)\right]\nonumber\\
&&+4f_{{\bf k}}g_{{\bf k}}\alpha_{{\bf k}}\beta_{{\bf k}}u_{{\bf k}}^+v_{{\bf k}}^+\left(\frac{1}{m^*_{{\bf k}i,j}}+\frac{1}{m^*_{{\bf k}+{\bf Q}i,j}}\right)
\Big(1-\Big\langle
B_{{\bf k},\uparrow}^{\dag}B_{{\bf k},\uparrow}\Big\rangle
-\Big\langle
B_{-{\bf k},\downarrow}^{\dag}B_{-{\bf k},\downarrow}\Big\rangle\Big)\nonumber\\
&&+f_{{\bf k}}^2\left(\left(\frac{1}{m^*_{i,j}}\right)\beta_{{\bf k}}^2
+\left(\frac{1}{m^*_{{\bf k}+{\bf Q}i,j}}\right)\alpha_{{\bf k}}^2\right)
\left[2(v_{{\bf k}}^+)^2+\Big((u_{{\bf k}}^+)^2-(v_{{\bf k}}^+)^2\Big)\right.\nonumber\\
&&~~~\left.\times\Big(1-\Big\langle
B_{{\bf k}+{\bf Q},\uparrow}^{\dag}B_{{\bf k}+{\bf Q},\uparrow}\Big\rangle
-\Big\langle
B_{-{\bf k}-{\bf Q},\downarrow}^{\dag}B_{-{\bf k}-{\bf Q},\downarrow}\Big\rangle\Big)\right]\nonumber\\
&&+g_{{\bf k}}^2\left(\left(\frac{1}{m^*_{i,j}}\right)\alpha_{{\bf k}}^2
+\left(\frac{1}{m^*_{{\bf k}+{\bf Q}i,j}}\right)\beta_{{\bf k}}^2\right)
\left[2(u_{{\bf k}}^+)^2-\Big((u_{{\bf k}}^+)^2-(v_{{\bf k}}^+)^2\Big)\right. \nonumber\\
&&~~~\left.\times \Big(1-\Big\langle
B_{{\bf k}+{\bf Q},\uparrow}^{\dag}B_{{\bf k}+{\bf Q},\uparrow}\Big\rangle
-\Big\langle
B_{-{\bf k}-{\bf Q},\downarrow}^{\dag}B_{-{\bf k}-{\bf Q},\downarrow}\Big\rangle\Big)\right]\nonumber\\
&&+4f_{{\bf k}}g_{{\bf k}}\alpha_{{\bf k}}\beta_{{\bf k}}u_{{\bf k}}^-v_{{\bf k}}^-\left(\frac{1}{m^*_{{\bf k}i,j}}+\frac{1}{m^*_{{\bf k}+{\bf Q}i,j}}\right)
\Big(1-\Big\langle
B_{{\bf k}+{\bf Q},\uparrow}^{\dag}B_{{\bf k}+{\bf Q},\uparrow}\Big\rangle
-\Big\langle
B_{-{\bf k}-{\bf Q},\downarrow}^{\dag}B_{-{\bf k}-{\bf Q},\downarrow}\Big\rangle\Big)\nonumber\\
&=&-\frac{e^2}{c\Omega}\sum_{j=1}^2a_j(0)\sum_{{\bf k}}^{\prime}\sum_{\nu=\pm}
\left(\frac{1}{M^{\nu}_{{\bf k}i,j}}\right)\left(1-\frac{E^{\pi}_{{\bf k}}+E^{m}_{{\bf k}}}{E^{\nu}_{{\bf k}}}\tanh{(\beta E^{\nu}_{{\bf k}}/2)}\right)
+\left(\frac{1}{m^{\nu}_{{\bf k}i,j}}\right)\tanh{(\beta E^{\nu}_{{\bf k}}/2)}.\nonumber\\
\end{eqnarray}
\end{widetext}


\begin{thebibliography}{99}
\bibitem{LSCO2000}A. Ino, C. Kim, M. Nakamura, T. Yoshida, T. Mizokawa, Z.-X. Shen, A. Fujimori, T. Kakeshita, H. Eisaki, and S. Uchida, 
Phys. Rev. B {\bf 62}, 4137 (2000).

\bibitem{ccoc} K. M. Shen, {\it et al.}, 
Phys. Rev. B {\bf 69}, 054503 (2004).

\bibitem{bi22122006}K. Tanaka, W. S. Lee, D. H. Lu, A. Fujimori, T. Fujii, Risdiana, I. Terasaki, D. J. Scalapino, T. P. Devereaux, Z. Hussain, and Z.-X. Shen, 
Science {\bf 314}, 1910 (2006).

\bibitem{bi2212} I. M. Vishik, {\it et al.} 
Proc. Nat. Acad. Sci. (USA) {\bf 109}, 18332 (2012).

\bibitem{LSCO2013}E. Razzoli, G. Drachuck, A. Keren, M. Radovic, N. C. Plumb, J. Chang, Y.-B. Huang, H. Ding, J. Mesot, and M. Shi, 
Phys. Rev. Lett. {\bf 110}, 047004 (2013).

\bibitem{bi2201} Y. Peng, J. Meng, D.Mou, J. He, L. Zhao, Y.Wu, G. Liu, X. Dong, S. He, J. Zhang, X. Wang, Q. Peng, Z. Wang, S. Zhang, F. Yang, C. Chen, Z. Xu, T. K. Lee, X. J. Zhou, 
Nat. Commun. {\bf 4}, 2459 (2013).

\bibitem{YBCO}D. Gustafsson, D. Golubev, M. FogelstrÃ¶m, T. Claeson, S. Kubatkin, T. Bauch, and F. Lombardi, 
Nat. Nanotechnology {\bf 8}, 25-30 (2013).

\bibitem{TanmoyED} T. Das, R. S. Markiewicz, and A. Bansil, 
Phys. Rev. Lett {\bf 98}, 197004 (2007).

\bibitem{CSTing}Qingshan Yuan, Xin-Zhong Yan, and C. S. Ting, Phys. Rev. B {\bf 74}, 214503 (2006).

\bibitem{TanmoyFeSe}T. Das, and A. V. Balatsky, 
Phys. Rev. B {\bf 84}, 014521 (2011).

\bibitem{NodelessFeSe1}T. A. Maier, S. Graser, P. J. Hirschfeld, and D.J. Scalapino, Phys. Rev.B, {\bf 83}, 100515 (2011).

\bibitem{NodelessFeSe2}Fa Wang, F. Yang, M. Gao, Z.-Yi Lu, T. Xiang, and D.-H. Lee, Europhys. Lett., {93}, 57003, (2011).

\bibitem{CoulombGap}W. Chen, G. Khaliullin, and O. P. Sushkov, 
Phys. Rev. B {\bf 80}, 094519 (2009).

\bibitem{Annica}A. M. Black-Schaffer, D. S. Golubev, T. Bauch, F. Lombardi, and M.Fogelstr\"om, Phys. Rev. Lett. {\bf 110}, 197001 (2013).

\bibitem{TSC}Y.-M. Lu, T. Xiang, and D.-H. Lee, 
arXiv:1311.5892.

\bibitem{CDMFT1}S. Sakai, and M. Civelli, 


\bibitem{CDMFT2}A. Go, and A. J. Millis, 
arXiv:1311.6819.

\bibitem{Tsg} C. Panagopoulos, J.L. Tallon, B.D. Rainford, T. Xiang, J.R. Cooper, and C.A. Scott, Phys. Rev. B {\bf 66}, 064501 (2002); M.-H. Julien, Phys. B Condens. Matter {\bf 329â€“333}, 693 (2003); S.-H. Baek, T. Loew, V. Hinkov, C.T. Lin, B. Keimer, B. BÃ¼chner, and H.-J. Grafe, Phys. Rev. B {\bf 86}, 220504 (2012); M. Enoki, M. Fujita, T. Nishizaki, S. Iikubo, D.K. Singh, S. Chang, J.M. Tranquada, and K. Yamada, Phys. Rev. Lett. {\bf 110}, 017004 (2013).


\bibitem{NMRBi11}S. Shimizu, S.-i. Tabata, S. Iwai, H. Mukuda, and Y. Kitaoka, 
Phys. Rev. B {\bf 85} , 024528 (2012).

\bibitem{NMRBi12}S. Shimizu, S.-i. Tabata, H. Mukuda, Y. Kitaoka, P. M. Shirage, H.i Kito, and A. Iyo, 
Phys. Rev. B {\bf 83} , 214514 (2011).

\bibitem{NMRTlHg12}H. Mukuda, S. Shimizu, A. Iyo, and Y. Kitaoka, 
J. Phys. Soc. Jpn. {\bf 81}, 011008 (2012).

\bibitem{NMRYBCO12}S.-H. Baek, T. Loew, V. H i n k o v, C. T. Lin, B. Keimer, B. B\'uchner, and H.-J. Grafe, 
Phys. Rev B {\bf 86}, 220504(R) (2012).

\bibitem{NeutronYBCO10}D. Haug, V. Hinkov, Y. Sidis, P. Bourges, N. B. Christensen, A. Ivanov, T. Keller, C. T. Lin and B. Keimer, New J. Phys. {\bf 12}, 105006 (2010) .

\bibitem{NeutronYBCO11}V. Bal\'edent, D. Haug, Y. Sidis, V. Hinkov, C. T. Lin, and P. Bourges,  Phys. Rev. B {\bf 83} , 104504 (2011)

\bibitem{twodomes} T. Das, C. Panagopoulos,  New J. Phys. {\bf 18}, 103033 (2016). 

\bibitem{FIRgap} S. Lupi, D. Nicoletti, O. Limaj, L. Baldassarre, M. Ortolani, S. Ono, Y. Ando, and P. Calvani, Phys. Rev. Lett. {\bf 102}, 206409 (2009).

\bibitem{dSCtriplet}G.â€‰C. Psaltakis and E.â€‰W. Fenton, J. Phys. C {\bf 16}, 3913 (1983); B. Kyung, Phys. Rev. B {\bf 62}, 9083 (2000).

\bibitem{so5}S.-C. Zhang, 
Science {\bf 275} 1089-1096 (1997).

\bibitem{cecoin5}A. Aperis, G. Varelogiannis, and P. B. Littlewood, 
Phys. Rev. Lett. {\bf 104}, 216403 (2010).


\bibitem{footnote} While a single band model is valid for the low-energy physics of most of the cuorates, other bands such as materials specific CuO chain state\cite{DasYBCO} or HgO like bands\cite{DasHg} are required to be included in YBCO and HgCCO cuprates, respectvely.

\bibitem{DasYBCO}T. Das, 
Phys. Rev. B {\bf 86}, 064527 (2012).

\bibitem{DasHg}T. Das, 
Phys. Rev. B {\bf 86}, 054518 (2012).

\bibitem{BobTB}R. S. Markiewicz, S. Sahrakorpi, M. Lindroos, Hsin Lin, and A. Bansil, 
Phys. Rev. B {\bf 72}, 054519 (2005).

\bibitem{Tanmoytwogap}T. Das, R.S. Markiewicz, and A. Bansil, 
Phys. Rev. B {\bf 77}, 134516 (2008).

\bibitem{LFu}L. Fu, and E. Berg, 
Phys. Rev. Lett. {\bf 105}, 097001 (2010).

\bibitem{so8}R. S. Markiewicz and M. T. Vaughn, 
Phys. Rev. B {\bf 57}, R14052 (1998).

\bibitem{kyung}B. Kyung, 
Phys. Rev. B, {\bf 62}, 9083 (2000).

\bibitem{Alloul}H. Alloul, J. Bobroff, M. Gabay, and P. J. Hirschfeld, Rev. Mod. Phys., {\bf 81}, 45–108 (2009); H. Alloul, P. Mendels, H. Casalta, J. F. Marucco, and J. Arabski, Phys. Rev. Lett. {\bf 67}, 3140–3143 (1991).

\end{thebibliography}
\end{document}